\documentclass[aps,prr,floatfix,twocolumn,superscriptaddress]{revtex4}
\usepackage{amsmath,amssymb}
\usepackage[usenames]{color}
\usepackage{graphicx}
\usepackage{epstopdf}
\usepackage{xspace}
\DeclareGraphicsExtensions{.eps}
\usepackage{mathbbol}

\newcommand{\be}{\begin{equation}}
\newcommand{\ee}{\end{equation}}
\newcommand{\bea}{\begin{eqnarray}}
\newcommand{\eea}{\end{eqnarray}}

\usepackage{rotating, graphicx}

\usepackage[T1]{fontenc}
\usepackage{booktabs}
\usepackage{threeparttable}
\newcommand{\tabincell}[2]{\begin{tabular}{@{}#1@{}}#2\end{tabular}}

\begin{document}	

%\begin{CJK}{GBK}{kai}
\title{Unsupervised learning of topological phase transitions using the Calinski-Harabaz index}
\author{Jielin Wang}
\affiliation{ College of information and computer, Taiyuan University of Technology, Taiyuan 030024, China}

\author{Wanzhou Zhang}
\thanks{zhangwanzhou@tyut.edu.cn}
\affiliation{
College of Physics and Optoelectronics, Taiyuan University of Technology, Taiyuan 030024, China}
\author{Tian Hua} %\footnote{co-first author: tianhua@tyut.edu.cn}}
\thanks{tianhua@tyut.edu.cn}

%\author{Tian Hua \footnote{co-first author: tianhua@tyut.edu.cn}}
\affiliation{ College of information and computer, Taiyuan University of Technology, Taiyuan 030024,  China}

\author{Tzu-Chieh Wei}
\thanks{tzu-chieh.wei@stonybrook.edu}
\affiliation{C.N. Yang Institute for Theoretical Physics and Department of Physics and Astronomy,
State University of New York at Stony Brook, Stony Brook, NY 11794-3840, USA}
\date{\today}
\begin{abstract}

Machine learning methods have been recently applied to learning phases of matter and transitions between them. Of particular interest is the topological phase transition, such as  in the XY model, whose critical points can be difficult to be obtained by using unsupervised learning such as the principal component analysis.
Recently, authors of [Nature Physics \textbf{15},790 (2019)] employed  the
   diffusion map method for identifying topological orders and  were able to determine the Berezinskii-Kosterlitz-Thouless (BKT) phase transition of the XY model, specifically via
   the intersection of the average cluster distance $\bar{D}$ and
the within-cluster dispersion parameter
$\bar\sigma$ (when the different clusters vary from separation to mixing together). However, sometimes it is not easy to find the intersection if
    $\bar{D}$ or  $\bar{\sigma}$ does not change too much due to topological constraint. In this work, we propose to use the Calinski-Harabaz ($ch$) index, defined roughly as the ratio  $\bar D/\bar \sigma$, to determine the critical points, at which the $ch$ index reaches a maximum or minimum value, or jumps sharply.

We examine the $ch$ index in  several statistical models, including ones that contain a BKT phase transition. For the Ising model, the peaks of  the quantity $ch$ or its components are consistent with the position of the specific heat  maximum. For the XY model, both on the square and honeycomb lattices,  our results of the $ch$ index  show the convergence of the peaks over a range of the parameters $\varepsilon/\varepsilon_0$ in the Gaussian kernel.

We also examine the generalized  XY model with $q=2$ and $q=8$ and  study the phase transition using the fractional $1/2$-vortex or $1/8$-vortex constraint  respectively.  The global phase diagram can be obtained by our method, which does not use the label of configuration needed by supervised learning, nor  a crossing from two curves $\bar{D}$ and $\bar{\sigma}$. Our method is thus useful to both topological and non-topological phase transitions and can achieve accuracy as good as supervised learning methods previously used in these models, and may be used for searching phases from experimental data.

\end{abstract}
\pacs{05.70.Fh, 64.60.ah, 64.60.De,64.60.at,89.20.Ff}
	%\pacs{75.10.Jm, 05.30.Jp, 03.75.Lm, 37.10.Jk}
\maketitle
%%%%%%%%%%%%%%%%%%%%%%%%%%%%%%%%%%%%%%%%%%%%%%%%%%%%%%%%%%%%%%%%%%%%%%%%
%%%%%%%%%%%%%%%%%%%%%%%%%%%%%%%%%%%%%%%%%%%%%%%%%%%%%%%%%%%%%%%%%%%%%%%%
%%%%%%%%%%%%%%%%%%%%%%%%%%%%%%%%%%%%%%%%%%%%%%%%%%%%%%%%%%%%%%%%%%%%%%%%
\section{introduction}
%{\it Introduction.---}
Exploring phases of matter and phase transitions is a traditional but still active research direction in statistical physics~\cite{pt,qpt}, partly due to new phases of matter that have been uncovered.
In recent years, this field of research is revived thanks to  the introduction of artificial intelligence and  machine learning methods  to recognize phases and transitions~\cite{roger1}.
Among the various methods, supervised learning methods are used to train the network using prior labeled data generated by Monte Carlo
methods in various classical systems, such as the Ising model and its variants~\cite{roger1,longising, smising}, XY models~\cite{hu,mlpxy,rogerxy,Wessel} with BKT phase transitions~\cite{xykt,bkt}, Dzyaloshinskii-Moriya ferromagnets~\cite{dm}, skyrmions~\cite{sky}, Potts models~\cite{Potts} and percolation  models~\cite{mlpxy}.  The properties of quantum systems, such as
the energy spectrum, entanglement  spectrum~\cite{confu1}, or configuration in the imaginary time~\cite{fermi,xuefeng} are also used as inputs to the neural networks for the training.
On the other hand,  unsupervised learning methods can provide unbiased  classification, as they do not need prior knowledge of the system so as
to classify phases and obtain the phase transition. These include methods such as the principal component analysis~\cite{wang1}, autoencoders~\cite{unp}, t-SNE~\cite{unsup1}, clustering with quantum mechanics  \cite{tony}. A  key feature of these is to determine the sought-after properties (such as phase transition) by examining indicators from the scattered plot in the reduced space.
Beyond equilibrium statistical physics, non-equilibrium and dynamical properties~\cite{dy,nonequi2,out2} are also obtained by machine learning methods.
In addition, there are other newly developed schemes applied to the phases of matter, such as the adversarial neural networks~\cite{adv}, confusion method and its extension~\cite{dcn}, the super resolving method~\cite{rs}, and even applications to experimental data~\cite{ex1, ex2}.
Other developments in this area can be found in Refs.~\cite{topo-inv,z2spin,tp2,mbl,jgliu,ml,selfmap,tool,spinor,conf,order,exp}
 and  the review article \cite{rev}.

It is well known that in traditional continuous phase transitions global symmetry is broken and these transitions can be
described by the Landau theory. However, the topological phase transitions have no
broken symmetry and therefore it is of great interest to understand how the transitions emerge and how to locate the transition
points. Recent developments of machine learning  offers new tools
for this endeavor \cite{hu,rogerxy,Wessel,mlpxy}, based
on supervised learning. However, the
learning-by-confusion scheme when applied to the XY model predicts a transition temperature set by the value located at the maximum of the specific heat~\cite{Wessel}.
In  Ref.~\cite{rogerxy}, it was found that significant feature engineering of the raw spin states is needed to relate vortex
configurations and the transition. Moreover, a single-hidden-layer
Fully Connected Networks (FCN) could not correctly learn the phases in the XY model, whereas the Convolutional Neural Networks (CNN) was successfully employed to learn the BKT transition \cite{rogerxy}, and the classification was later extended to the generalized XY model~\cite{mlpxy}. However, for unsupervised learning with the PCA method, it was claimed to be difficult to identify the transition~\cite{hu}.
Recently, Rodriguez-Nieva and  Scheurer used the  diffusion map method~\cite{nature} invented by R.R. Coifman and S. Lafon~\cite{lafon} and related to quantum clustering~\cite{tony}, and devised an unsupervised learning method
for identifying topological orders and successfully locating the BKT transition.
They divided the configurations into several topological sectors with different winding numbers, then calculated the eigenvector $\Psi$ and eigenvalues $\lambda$  of the so-called diffusion matrix $P$ (see Sec.~\ref{sec:dif}). From the intersections of  the average
cluster distance $\bar{D}$ and within-cluster dispersion $\bar\sigma$,   or equivalently the intersection of $\Delta\lambda$ (the jump in the eigenvalues) and $\sigma_{\lambda}$ (the standard deviation of the eigenvalues), the phase transition
of the XY model on the square lattice can be obtained very well.

Our motivation of this study is to examine whether or not the diffusion map (DM) method of Rodriguez-Nieva and  Scheurer (RNS method)
is suitable beyond the pure XY model, such as the generalized XY model. Indeed we find that the DM method works for some topological phase transitions, but it fails to locate the phase transition in the generalized XY model in other regimes.
%Whether the method could apply to the XY model on other lattices such as, the honeycomb lattices
%having a different phase transition point.
Specifically, the RNS method for determining the transition point relies on the intersection of the two curves (such as the average
cluster distance $\bar{D}$ and within-cluster dispersion $\bar\sigma$), and in the $q=8$ generalized XY model, as illustrated in Sec.~\ref{sec:gxy}, we cannot find an
intersection there.  The thermal fluctuation
% TCS: make a new sentence
is not strong enough and the scattering clusters with different
winding as numbers do not mix close to
 $T_c$, i.e. $\bar{D}$  does not decrease substantially.
The question arises: are there  other quantities that can be used to locate the phase transition points?

In this work, we mainly  use the Calinski-Harabaz (ch) index score~\cite{CH} defined by  $ch_t/ch_b$, related to the ratio of  $\bar D/\bar \sigma$, which means that if the variation of $\bar {D}$ can be negligible due to  strong topological constraints, the variation of $\bar \sigma$
itself can help to determine the phase transition point. We also use
the Silhouette coefficient ($sc$)~\cite{lunkuo} or its components  to  compare with the results.

The outline of this work is as follows. Sec. \ref{sec:method} shows
the DM methods, the definition of the indices $ch$, $sc$ and their
 components, their advantage and prior knowledge for using the indices. Sec. \ref{sec:ising}  shows the signature  of $ch$ and $sc$ for the two-dimensional Ising model   from configurations  using the Swendsen-Wang algorithms \cite{SW}.
In Sec. \ref{sec:2dxy}, the critical phase transition points of the XY model
on the square and the honeycomb lattices are obtained using the DM method.
In Sec.~\ref{sec:gxy}, for the $q=2$ and $q=8$ generalized XY models,  the global phase diagrams are obtained
by the DM method assisted by PCA or k-PCA methods.
Other techniques are discussed in Sec.~\ref{sec:otm} regarding the effect of
higher dimensions considered in the k-means method and how to automatically find the hyperparameter $\varepsilon/\varepsilon_0$.
Concluding comments are made in Sec.\ref{sec:con}.
In Appendix \ref{sec:1dxy}, the simplest example  1D XY
model is discussed, and the comparison between PCA and k-PCA is shown in Appendix \ref{sec:pca}.
Finally,  a total of 11 indices   used in unsupervised learning are listed in Appendix \ref{sec:11ind}.

\section{Methods}
\label{sec:method}
\subsection{The review of  diffusion map method}
\label{sec:dif}
Here, we explain the DM method of Rodriguez-Nieva and  Scheurer~\cite{nature}. Assume that we have   $M$ configurations  $\{x_l\}$, where $l=1,\dots M$.
The connectivity  between $x_l$ and $x_{l^{'}}$ is denoted by the elementary Gaussian kernel
\begin{equation}
	\label{eq:k}
	K_{\varepsilon}(x_{l},x_{l^{'}})= \exp (-\frac{||{x_{l}-x_{l^{'}}}||^{2}}{2N\varepsilon}).
\end{equation}
The normalization  of $K_{\varepsilon}(x_{l},x_{l^{'}})$ can be done by performing the sum over $l^{'}$,
\begin{equation}
\label{eq:p}
	P_{l,l^{'}}= \frac{K_{\varepsilon}(x_{l},x_{l^{'}})}{z_{l}},z_{l}=\sum^{m}_{l^{'}=1}K_{\varepsilon}(x_{l},x_{l^{'}})
\end{equation}
We  can also perform the normalization along the direction of $l$ (i.e., the sum over $l$).
%Then value of $P_{l,l^{'}}$ is $\Phi_{l}$.
The eigenvalue equation of the diffusion matrix $P_{l,l^{'}}$ is $P\psi_k=\lambda_k\psi_k$, where
 $\psi_k$'s are the right eigenvectors, with
the corresponding eigenvalues
$\lambda_{k}\leq 1$, for
$k=0,1,\dots,m-1$.
%\begin{equation}
%\label{Eql}
%	l \rightarrow \Phi _{l} : = [(\psi )_{1} ) _{l} , (\psi)_{2} ) _{l} , \ldots, (\psi )_{m-1} ) _{l} ]
%\end{equation}

To find the phase transition point, Ref.~\cite{nature} and its earlier preprint version \cite{prep} use two different ways, respectively:
\renewcommand{\theenumi}{\alph{enumi}}%
%\begin{enumerate}
\\
(a)
Intersection of the mean distance of cluster centers $\overline{D}/2n$, where $n$ is the number of clusters, and the dispersion  $\overline{\sigma}$ of the data
points in each cluster. The quantities $\overline{D}/2n$ and  $\overline{\sigma}$ are obtained from the scatter plot, where $n$ is the number of topological sectors.% ({\bc define $\overline{D}$; $\overline{\sigma}$ is ?}).

After obtaining the eigenvectors of the $P$ matrix given by
\begin{equation}
\label{Eql}
	\Phi : = [(\psi )_{1}  , (\psi)_{2}  , \ldots , (\psi )_{m-1}  ],
\end{equation}
the authors project them unto a two-dimensional space, namely, a two-column matrix, then $\overline{D}/2n$ and the dispersion  $\overline{\sigma}$
can be
obtained from the scatter plot of the two column vectors or their modified version. The detailed application to
the one-dimensional XY model is reproduced in Appendix \ref{sec:1dxy}.
%\item
\\
(b) Intersection of  the mean fluctuation  $\sigma_{\lambda}$ and gap of eigenvalues  $\Delta\lambda$, where
\begin{equation}
\sigma_{\lambda} = \frac{1}{n}\sum^{n-1}_{k=0}(\lambda_k-\overline{\lambda})^2 \;,\; \overline{\lambda}=\frac{1}{n}\sum_{k=0}^{n-1}\lambda_{k},
\label{eqsigma}
\end{equation}
and the gap of eigenvalues between the  topological sectors $n$ and $n-1$,
\begin{equation}
\Delta\lambda = \lambda_n - \lambda_{n-1}.
\label{eqdelta}
\end{equation}
%\end{enumerate}

%The Silhouette coefficient  proposed by Peter J. Rousseeuw\cite{lunkuo} in the year 1986 , which quantifies the quality of clustering achieved.
%validates the clustering
%performance based on the pairwise difference of between-
%and within-cluster distances.
%When the cluster could not be distinguished, the Sc will begin decrease from its maxvalue 1.

%
%
%The \tony{work} is arranged as follows.
%In section \ref{sec:dif} describe the algorithms and measured quantities that can
%obtain the phase points. A example of one dimension is also present.
%In section \ref{sec:mod-res}, models and results are present.

%%%%%%%%%%%%%%%%%%%%%%%%%%%%%%%%%%%%%%%%%%%%%%%%%%%%%%%%%%%%%%%%%%%%%%%%
%%%%%%%%%%%%%%%%%%%%%%%%%%%%%%%%%%%%%%%%%%%%%%%%%%%%%%%%%%%%%%%%%%%%%%%%
%%%%%%%%%%%%%%%%%%%%%%%%%%%%%%%%%%%%%%%%%%%%%%%%%%%%%%%%%%%%%%%%%%%%%%%%
\subsection{The indices \textbf{\textit{ch}} and \textbf{\textit{sc}}}

 %We first use  $ch$ index score when  applying  k-means to the reduced feature space $\varphi_1$ and $\varphi_2$ ($\varphi_l=[(\psi_1)_l,(\psi_2)_l,....,(\psi_k)_l]$,$k=1,2,...,n-1$),
% and then use the $ch$ index for the result of k-means.
We propose to use indices instead of intersections.
Based on the first two leading eigenvectors  $\Psi_0$ and $\Psi_1$ of the PCA, kernel-PCA (kPCA) or the second and third vectors $\Psi_1$ and $\Psi_2$ of the DM method, for a manually chosen cluster numbers $k$,  the $ch$ index is given
in terms of the following ratio:% of the between-clusters dispersion mean and the within-cluster dispersion:
\be
ch=\frac{ch_t}{ch_b}=\frac{tr(B_k)}{tr(W_k)}\times\frac{N-k}{k-1},
\label{eq:ch}
\ee
where $B_k$ is the between-group dispersion matrix and $W_k$ is the within-cluster dispersion matrix  and they are  defined as follows,
\be
\begin{array}{l}
W_k = \sum_{q=1}^k \sum_{x \in C_q} (x - c_q) (x - c_q)^T,  \\
B_k = \sum_q n_q (c_q - c) (c_q - c)^T,
\end{array}
\ee
where $N$ is the number of data points,
$C_q$ is the set of points in cluster $q$,
$c_q$ is the center of cluster $q$, and $c$ denotes the average center of all cluster centers $\{c_q \}$, and $n_q$ the number of points in cluster $q$.
The $sc$ index of the $i$-th sample is
\be
sc(i)=\frac{b(i)-a(i)}{max(b(i),a(i))}
\label{eq:sc}
\ee
where $a(i)$ is the mean distance between sample $i$ and all other data points in the same cluster, $b(i)$ is the mean dissimilarity of point $i$ to some cluster $C$ expressed as the mean of the distance from $i$ to all points in $C$ (where  $C\neq C_{i}$).
The mean $sc=\sum sc(i)/N$ over all points of a cluster is a measure of how tightly grouped all the points in the cluster are.
Sometimes $sc_a=\sum a(i) /N$ or $sc_b=\sum b(i)/N$ are also very useful \cite{11index}.

The performance of a total of 11 indices is shown in Appendix \ref{sec:11ind} and the results of indices $ch$, $dn$, $pbm$ and $Ii$ applied to the $q=8$ generalized XY model are all reasonable choices. Here, we choose two representative indices $ch$ and $sc$ as examples.
\begin{figure}
	\includegraphics[width=0.8\linewidth,height=0.8\linewidth]{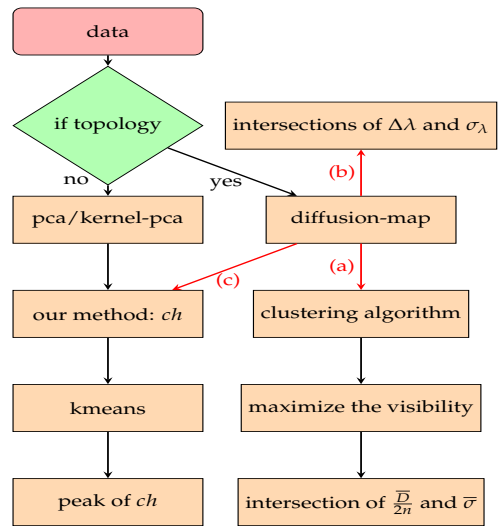}
	\caption{(Color online) Flow chart and main steps of the RNS method  and our modifications. The main difference begins at how to using the DM method (a) clustering algorithm in RNS method (b) intersections of $\Delta \lambda$ and $\sigma_{\lambda}$ (c)
Our method using $ch$.}
	\label{fig:flow}
\end{figure}

%
%\textcolor[rgb]{1.00,0.00,0.00}{B1:}\\
%In other words: How really unsupervised is this approach? Can
%the authors give at least a general comment on how much prior
%knowledge required for such methods to work for Hamiltonians as
%complicated as GXY?\\
%
%of  Silhouette index will be highest and kept at 1 for $T<Tc$,
%TCS: Tc or T_c?
%if we input correct the number of  clusters $n$,  equals to number of topological constraints, or winding number)  initialized before.
% When  $T>Tc$ the  Silhouette index begin to decrease.
% But the method has limitations will discussed.
%

\subsection{ Advantage and prior
knowledge required using  \textbf{\textit{ch}}}
\label{sec:adv-pri}

%\textcolor[rgb]{0.00,0.07,1.00}{A1:}
Fig.~\ref{fig:flow} draws the flowchart of the RNS method
and the place of our modifications.
Depending on whether or not we are considering topology,
we perform dimensional reduction by  DM or PCA (k-PCA) respectively, to get the %\textcolor[rgb]{1.00,0.00,0.00}
features of the data, i.e., the eigenvectors of  the diffusion matrix $P$ or the covariance matrix.
%\tony{Note: DM involves the diffusion matrix - is the order right?}.
Then we apply k-means to cluster the two-dimensional or
higher dimensional scattering points. If we do not choose ``ch'', then intersections of $\Delta\lambda$ and $\sigma_{\lambda}$,  or intersections of $\bar{D}/2n$ and $\bar{\sigma}$
 \cite{nature,prep} can be used instead.
%, {\color{red} which are listed in the flow chart in  Fig.~\ref{fig:flow} }.

To test whether the clustering is good or bad, if  ``$ch$'' is chosen, then
the peaks of $ch$ or its components $ch_t$ and $ch_b$ will be used
directly.
For the topological phase transition, the index $ch$ or its component
$ch_t$ and $ch_b$ can give signatures of phase transition when it is not easy to determine the  intersection
of the RNS method or when there is no intersection.

The index $ch$ can also be applied to non-topological phase transitions, such as
the order-disorder phase transition of the Ising model. This does not require  any prior knowledge except for the configurations generated  by e.g. Monte Carlo methods or from real experiments.

%\textcolor[rgb]{1.00,0.00,0.00}{B1:}
For topological phase transitions, such as the XY and generalized
XY models, although this type of unsupervised learning is not similar to supervised learning
(such as fully-connected layers, or convolutional neural
networks), it still needs labels of  configurations, i.e., the topological winding number and the number of possible phases.
However, the label of the topological winding number does not mean the label of phases,
 and essentially, this
method is still an unsupervised learning method.
%%%%%%%%%%%%%%%%%%%%%%%%%%%%%%%%%%%%%%%%%%p%%%%%%%%%%%:%%%%%%%%%%%%%%%%%%%
%%%%%%%%%%%%%%%%%%%%%%%%%%%%%%%%%%%%%%%%%%%%%%%%%%%%%%%%%%%%%%%%%%%%%%%%
%%%%%%%%%%%%%%%%%%%%%%%%%%%%%%%%%%%%%%%%%%%%%%%%%%%%%%%%%%%%%%%%%%%%%%%%
%{\it{2D Ising model.}---}
%\section{2D classical models}
\section{The 2D Ising model}
\label{sec:ising}
To test the ability to locate  the phase transition point $T_c$,
we calculate  $ch$ for the simplest model, i.e., the Ising model,
\be
H=-J\sum_{\langle i,j\rangle}S_iS_j,
\ee
where $J$ is the ferromagnetic interaction between a
pair of nearest neighborhood spins, and $S_i=\pm 1$. The unsupervised
learning of Ising model has been studied before, (see e.g., Refs.~\cite{wang1, unp}).
%All spins points up or down because the configurations {$\sigma$} obeys the Boltzman distribution
%$e^{-E_{\sigma}/kT}$ at $T<T_c$, where  the status of spin directions will become disordered when $T>T_c$.

%\begin{figure}
%	\includegraphics[width=0.90\linewidth]{fig2a-f.eps}
%\vskip 0.5 cm
	%\hskip 0.5 cm
%		\includegraphics[width=0.9\linewidth]{fig2g-i.eps}

%\label{fig:is}
%\end{figure}
%Here, we show how to use $ch$ and $sc$ to determine $T_c$ shortly.
We use a  two-dimensional $64\times 64$ lattice and generate
$N_s=2000$ samples  for each temperature $T_i$ and
analyze them by PCA, using  the scattering data of the first two leading eigenvectors $\Psi_0$ and $\Psi_1$.
Moreover, we calculate $ch$($k=2$) and sc($k=2$) for each $T_i$
according to Eqs.~(\ref{eq:ch}) and (\ref{eq:sc}).

\begin{figure}[htb]
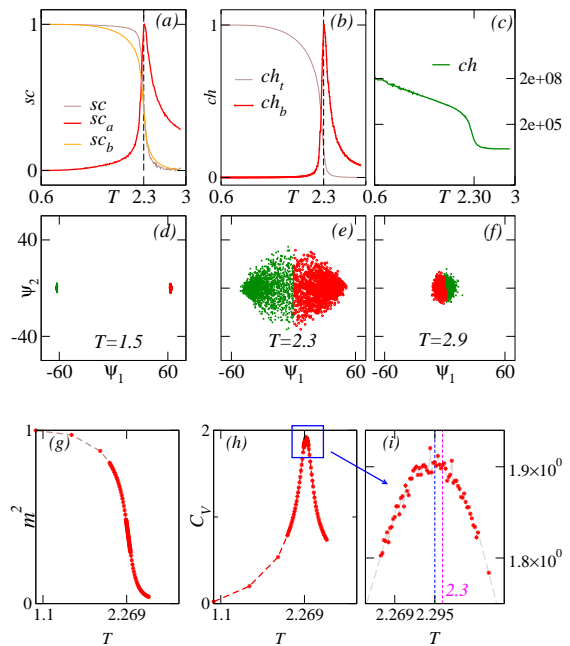

	\includegraphics[width=0.85\linewidth]{fig3-wang.eps}
\vskip 0.5 cm
	\includegraphics[width=0.85\linewidth]{fig3-i-copy.eps}

\caption{ For the Ising model with sizes $L=64$, (a)   $sc$, $sc_a$, $sc_b$ (b) $ch_t$ and $ch_b$ presented in
the normalized  range $[0,1]$  (c) $ch$ (d) - (f)  scatter plot with  $\Psi_1$ and $\Psi_2$,
 at three typical temperatures $T< T_c$, $T \approx T_c$ and $T > T_c$, respectively, where different colors mean labels generated by k-means.
(g)-(i) Monte Carlo results $m^2$, $C_v$ and zoomed $C_v$. The peaks are at 2.295, consistent with 2.30.}
\label{fig:is}
\end{figure}

Figure~\ref{fig:is} shows the main results for the Ising model. In Fig.~\ref{fig:is} (a), $sc$ itself  and $sc_b$ have a sharp decrease whereas $sc_a$
has peaks around $T_c$ (here we re-scale the results so the maximum value is 1).
In Figs.~\ref{fig:is} (b) and (c),
the $ch_b$ index is peaked around $2.3$;
$ch_t$ and $ch$  also have a sharp jump  at the phase transition points.

To understand these results,  the scatter plot of $\Psi_1$ and $\Psi_2$ are shown in Figs.~\ref{fig:is} (d)-(f) for temperatures $T=1.5$, 2.3 and 2.9, respectively.
 At low
 temperature, the clusters identified by the  two colors separate from each other   in the reduced space and finally
 mix together at high temperatures.

In general, the $ch$ index is
large when the clusters are well separated and the points in each
cluster are well aggregated. From this viewpoint, in the low temperature phase,
the $ch$ index becomes large because  all up states and
all down states can be easily distinguished if the analysis is
successful.

The index $ch$ and their components  % with ($k=2$, assuming there are two types of data)
perform well in detecting the transition, and the position of the peak or the jump  is the largest  at $T=2.3$.
Around
the phase transition point, configurations
possess the properties from both  phases (paramagnetic and  ferromagnetic)
and the fluctuation is the largest there.

Comparing to traditional Monte Carlo results with
the same size $L=64$ as shown in Fig.\ref{fig:is} (g)-(i),
we find that the position of the peak is located at  %\tony{peaks of the specific heat is/are?  peaks is plural. Just what does have this value at 2.295?}
around 2.295 consistent with our $ch$ or $sc$ results with numerical intervals of 0.01.
For the purpose of reference, the thermodynamic limit transition $T_c =2.269$ is marked.

\vskip 1cm

%\noindent \textcolor[rgb]{0.00,0.07,1.00}{A3}:
It should be noted that if we simulate the Ising model by the
Metropolis Monte Carlo algorithm \cite{metro}, which flips  one spin each time, in  the low temperature limit,  almost all spins  choose  to sit in   the initial state (i.e., $111111$, spin up). The scatter plot will simply have one group.
 However, this is not wrong because the state still obeys the Boltzmann distribution.  This is the reason of small $ch$ at low temperatures
%\tony{when? - ``of'' is not right.}
of using the Metropolis algorithm to generate configurations. However, we can still observe the signature at the phase transition point.
Using the Swendsen-Wang algorithm instead \cite{SW}, i.e., a  global updating Monte Carlo method, the spin states will all be spin up or spin down in lower temperatures, which is not dependent of the initial state.

%%%%%%%%%%%%%%%%%%%%%%%%%%%%%%%%%%%%%%%%%%%%%%%%%%%%%%%%%%%%%%%%%
%{\it 2D XY model.--}
\section{The 2D XY model}
\label{sec:2dxy}
The Hamiltonian of the classical XY model is given by
\begin{equation*}
H  = -J\sum\cos(\theta_i-\theta_j),
\end{equation*}
where $\left \langle i,j \right \rangle$ denotes a nearest-neighbor pair of sites $i$ and $j$, and
$\theta$  in $(0, 2\pi]$ is a classical variable defined at each site   describing the angles of spin directions in a two-dimensional spin plane.  The
sum in the Hamiltonian is over nearest-neighbor pairs on the square lattice $(L\times L)$ with the periodic
boundary condition; other lattices can be also considered.
%The examples of the XY model on the square and honeycomb lattices are discussed in the following.
\subsection{The 2D XY model on square lattices}
Now we analyze the first example, i.e., the two-dimensional XY model on the square lattice.
Firstly, we generate the configurations with five fixed winding number pairs
at $\nu=(\nu_x,\nu_y)=(0,0)$, $(0,1)$, $(1,0)$, $(0,-1)$ and $(-1,0)$.
For each fixed winding number pair,
it should be noted that, for the two-dimensional geometry, the winding number component $\nu_x=1$ means that the spins in each row form a winding number of $1$ rather than
just the spins on one  row randomly selected. %\tony{Badly explained.  Is a winding number a pair of numbers?  If so, how can it be one?}%({\bc check?}).
Cluster simulation algorithms~\cite{wf,SW} are not suitable to update the spins because
the global flips  break the topological winding number easily and therefore the Metropolis algorithm  is used while trying
to rotate the spin vector with a very small step each time so as to preserve the winding number.% If
%the random()$<e^{-\Delta E/K_BT}$ and the new configuration does not break the winding number
%in the correspoinding row and column.

%The way initialize a configuration the configuration like the way for the one-dimensional chain, illustrated in the appendix.
%Between different rows,  $\overline{\theta}^{(r)}$ is a little different about $2\pi/L$, very similar to 1D between-samples assigned $\theta^{(l)}$.
%
%In Fig.\ref{fig:xymodelafterevolution}, we  show a typical spin for $(\nu_x,\nu_y)= (0, 1) $ and the {\bc{$temperature= ?$. xxx times of monte carlo steps of updating.
%The spins along $y$ (vertical) direction have a rotationing with angles $2\pi$ }}

\begin{figure}[htb]
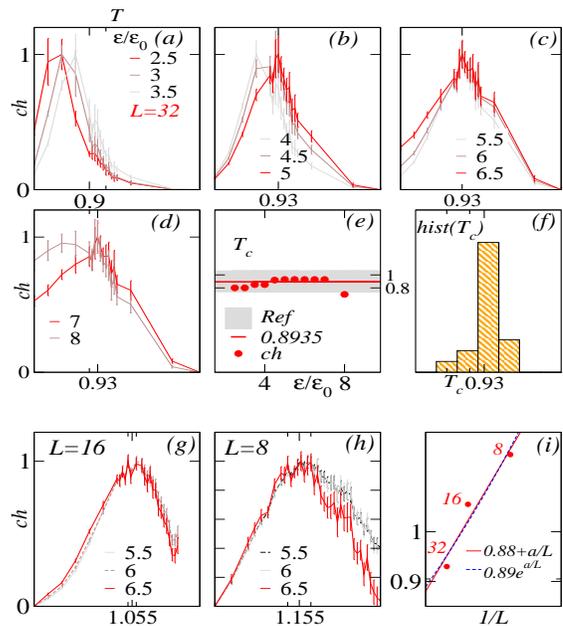

\includegraphics[width=0.85\linewidth,height=0.6\linewidth]{fig4-chss.eps}
\vskip 0.4cm
\includegraphics[width=0.85\linewidth,height=0.3\linewidth]{3chss.eps}

 \caption{ (Color online) $ch$ index for the XY model on two-dimensional square lattice of size $L=32$,
  with various value of $\varepsilon/\varepsilon_0$  (a) $2.5$, $3$, $3.5$   and
  (b) $4$, $4.5$, $5$   (c) $5.5$, $6$, $6.5$ (d)  $7$, $8$, $9$.
The error bar is obtained
  by (e) comparing the $T_c$ vs. $\varepsilon/\varepsilon_0$ using $ch$, Ref.~\cite{nature}, and the latest MC result \cite{kao}. (f) Histogram of $T_c$ obtained by using various $\varepsilon/\varepsilon_0$.
  (g)-(i) $ch$ index for the XY model on two-dimensional square lattice of size $L=8$ and $L=16$, and
     the scaling of estimated critical points.
  }
        \label{fig:chi}
\end{figure}

For each topological sector, we generate  $m=500$ configurations. Combining all configurations $\{x_l, l=1,\cdots,5m\}$ from all five sectors, we construct the kernel $K_\varepsilon$.
The elements between $K_{\varepsilon}(x_{l},x_{l^{'}})$ is defined in Eq.~(\ref{eq:k}) and the normalized matrix
$P_{\varepsilon}(x_{l},x_{l^{'}})$ is obtained, obeying its eigenvalue equation
$P\Psi_k=\lambda_k\Psi_k$.  Using the scatter plot of the second and third
leading eigenvectors $\Psi_1$ and $\Psi_2$ of $P_\varepsilon$, the $ch$ index are obtained and displayed in  Figs.~\ref{fig:chi} (a)- (d).
There is a parameter $\varepsilon$ in the above matrices and in order to deduce consistent results, we need to make sure the results are converging for a finite range of $\varepsilon/\varepsilon_0$. We see that this is indeed the case for  different values of $\varepsilon/\varepsilon_0$ in Figs.~\ref{fig:chi}(a)-(d).
For  $\varepsilon/\varepsilon_0 = 2.5$, $3$, $3.5$ \& $4$, $T_c$ is less than
$0.9$,   and then becomes  $0.93$  when $\varepsilon/\varepsilon_0 =  4.5$, $ 5$, $ 5.5$, $ 6$, $6.5$ and $7$.
Finally, the peaks move left and deviate from $0.93$ again
 when $\varepsilon/\varepsilon_0 = 8$ or larger.
 It should be noted that here, $10,000$  or more Monte Carlo steps, are used in order to reach the equilibrium of systems.

 The estimated points are labeled by
 the circles in Fig.~\ref{fig:chi} (e) and they all distribute nearby or on the red lines representing the latest result critical temperature $T_c=0.8935$\cite{kao}% - what is this number?
%Temperature? Can you give its units? } \cite{kao}.
 The intersections of  $\bar{D}$ and $\bar{\sigma}$ are labeled  by the gray regions in the critical regimes $T_c= 0.9\pm 0.1$  from Ref.~\cite{nature}.  It appears that  it is easier to use $ch$ to locate the phase transition as we only need to identify the peak location. The results from Ref.~\cite{nature} have greater uncertainty than those by using the $ch$ index.  %We also don't need to maximize the visibility of the clusters, as proposed in Ref.~\cite{nature}.

The histogram of our estimated  $T_c$ is shown in Fig.~\ref{fig:chi} (f), which helps to determine the hyperparameter $\varepsilon/\varepsilon_0$
(see Sec.~\ref{sec:otm}).

In Fig.~\ref{fig:chi} (g)-(i), the finite size effect of $T_c$ is checked using the $ch$ with two smaller sizes $L=8$ and $L=16$ with $\varepsilon/\varepsilon_0=5.5$, $6$ \& $6.5$. Combining the estimated $T_c (L)$ with $L=8, 16, 32$, we use  two different
ways of (linear, exponential) extrapolation to get $T_c(\infty)$ in the thermodynamic limit. The results are $0.88(5)$ and $0.89(5)$, respectively.
%
%In Figs.~\ref {fig:jdb}, we use the method of find  intersections
%of $\Delta \lambda_k$ and $\sigma_k$ for different $\varepsilon/\varepsilon_0$.In the whole range of $0<\varepsilon/\varepsilon_0<10$,
%
%Maybe considering more fluctuations and keeping the topological constraint simutaneously will help to  get more reasonable intersections. \\
\subsection{2D XY model on honeycomb lattices}
Here, we study the second example , i.e.,  the pure XY model on the honeycomb lattice,
as the critical point is known exactly at
 $T_c=1/\sqrt{2}\approx 0.707$ \cite{0.7}.

\begin{figure}[pbt!]
\includegraphics[width= \linewidth,height=0.6\linewidth]{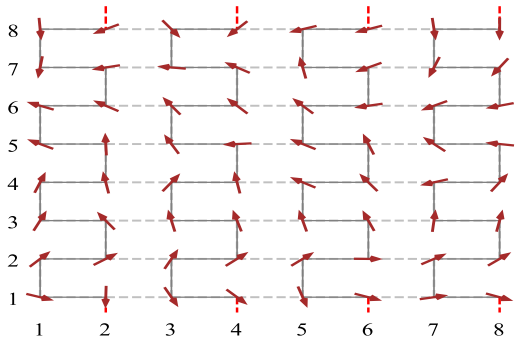}
	\caption{(color online) A $8\times 8$ honeycomb lattice equivalent to a brick-wall lattice.
 The gray solid line connects the spin forming the winding number in the $y$-direction, i.e. ($\nu_x$, $\nu_y$)= (1, 0). }
\label{fig:brickwall}
\end{figure}

The geometry of the honeycomb lattice is equivalent to  the $8 \times 8$ brick-wall lattice shown in Fig.~\ref{fig:brickwall}, where every spin
has three nearest-neighbor spins.
To initialize the configurations with a fixed winding number ($\nu_x$, $\nu_y$)=(0, 1),
the spins connected by solid gray  lines are defined as forming $\nu_y=1$ in the vertical direction.
Specifically,
if we start a spin  at position (2,1), and then go left to    $\rightarrow$ (1,1) $\rightarrow$ (1,2) $\cdots$  (2,8) and go back to  (2,1) through the red dashed lines, connecting the sites at the boundaries for periodic boundary conditions, the spins sweep an angle of $2\pi$ counter-clockwise.

 It should be noted that two spins, such as (1,1) and (2,1), connected by the horizontal gray lines will contribute to the winding number $\nu_y$, and they also contribute to $\nu_x$. This poses a problem when fixing $\nu_x$ and $\nu_y$ independently.
 Fortunately, this problem can be solved.
%the contributions of the difference between \tony{a pair nearest spins in the horizontal spins? - makes no sense!  What do you mean?}  will cancel each other.
Specifically, in the first row, labeled by 1 in the vertical ($y$) direction, the relation of angles  obeys
$\theta_{(2,1)}-\theta_{(1,1)}   = - (\theta_{(1,1)}-\theta_{(2,1)}) \approx - (\theta_{(3,1)}-\theta_{(2,1)})$.
During the simulation,  small fluctuations are allowed if they do not break
the winding number.

\begin{figure}
\includegraphics[width= \linewidth,height=0.6\linewidth]{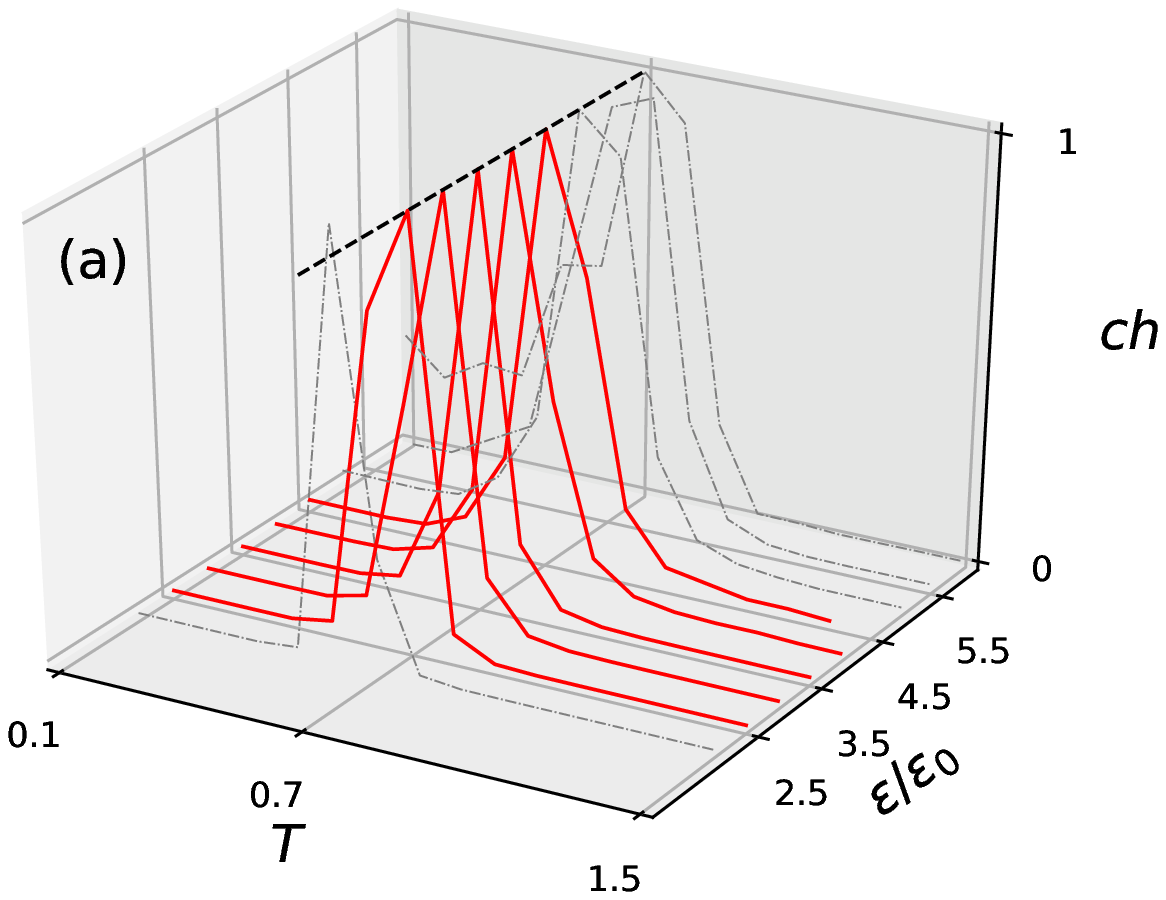}
\vskip 0.5cm
\includegraphics[width= 0.7\linewidth,height=0.4\linewidth]{new_sixHoney.eps}

\caption{(Color online) Locating the phase transition points of the pure XY model on
the honeycomb lattices (a) the intersections by $\Delta \lambda$, $\sigma_{\lambda}$
(b) $ch$ index (c) Comparison results between (a) and (b) and the exact result $\sqrt{2}/2$. }
	\label{fig:6res}
\end{figure}

%%%%%%%%%%%%%%%%%
	In Fig.~\ref{fig:6res},  using configurations constrained in five topological sectors on the $32\times 32 $ lattices, we find  that
the peaks are located at the exact value $T_c=\frac{1}{\sqrt{2}}\approx 0.7$ for several different values
of $\varepsilon/\varepsilon_0$ in the interval [2,6] with intervals of 0.5.
We also calculate the
intersections by  $\Delta \lambda$ and $\sigma_{\lambda}$ at
$\varepsilon/\varepsilon_0=$  5, 6, 7.
The intersection can also arrive  at $0.7$ when $\varepsilon/\varepsilon_0=7$, but not 5 and 6.
It indicates that the $ch$ performs better than the intersection as the intersection method may not give a high confidence of the transition.

\vskip 0.5cm

\begin{figure}[tbp!]
\includegraphics[width=0.95\linewidth]{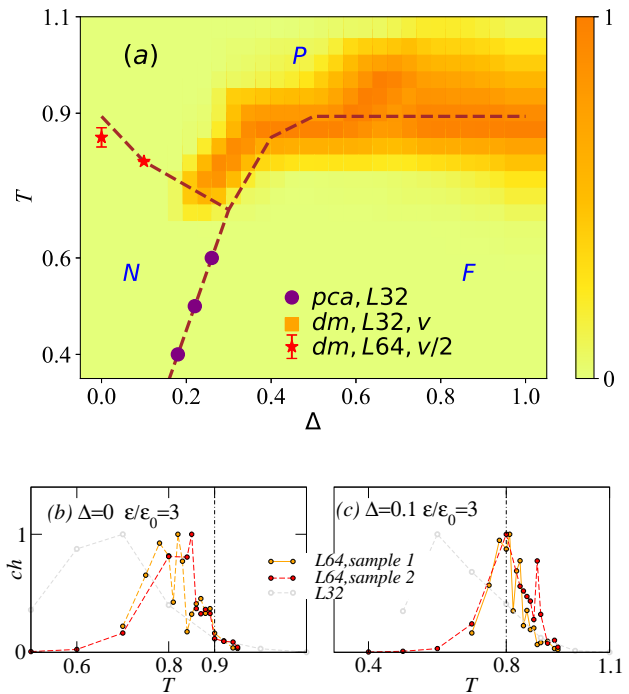}
\vskip 0.5 cm
\includegraphics[width=0.95\linewidth]{fig7b.eps}
\caption{(Color online) (a) The phase diagram of the $q=2$ GXY model, containing  $N$, $F$ and  $P$, by calculating the $ch$ index.
The dashed lines are from Ref.~\cite{q2mc}
(b) $N$-$P$ transition at fixed $\Delta=0$ : $ch$ vs. $T$. (c) $N$-$P$ transition $\Delta=0.1$:  $ch$ vs. $T$. The peaks are closer to the known values using the half-vortex constraint with $L$ up to $64$ when compared with those using only integer vortices.  }
\label{q2ph}
\end{figure}

%%%%%%%%%%%%%%%%%%%%%%%%%%%%%%%%%%%%%%%%%%%%%%%%%%%%%%%%%%%%%%%%%%%%%%%%%%%
%%%%%%%%%%%%%%%%%%%%%%%%%%%%%%%%%%%%%%%%%%%%%%%%%%%%%%%%%%%%%%%%%%%%%%%%%%%%
%%%%%%%%%%%%%%%%%%%%%%%%%%%%%%%%%%%%%%%%%%%%%%%%%%%%%%%%%%%%%%%%%%%%%%%%%%%%%
\section{2D Generalized XY (GXY) models }
\label{sec:gxy}

Here, we apply our method to the generalized XY models \cite{q8mc,q2mc}, whose Hamiltonian is given by
\begin{equation*}
H = -\sum[\Delta\cos(\theta_i-\theta_j )+(1-\Delta)\cos(q\theta_i-\theta_j)],
\end{equation*}
where $\Delta$ is the relative weight of the pure $XY$ model,
and $q$ is another integer parameter that can drive the
system to form a nematic phase. For both $\Delta=0$ and $\Delta = 1$
the model reduces to the pure XY model (redefining $q$ as
$1$ in the first case), and hence the transition temperature
is identical to that of the pure XY model. However, such
a redefinition is not possible with $\Delta \neq  1$.
The phase diagrams of the GXY models  
depend on the integer parameter $q$, and they have been explored extensively.
\subsection{\textbf{\textit{q=2}} GXY model}

The  $q=2$ GXY model has a richer phase diagram than the XY model and has an additional new phase~\cite{mlpxy}.
Thus, it is a good candidate model to test our method away from the pure XY limit.
The phase diagram is illustrated in Fig.~\ref{q2ph} (a)
on
the $\Delta-T$ plane, and we also show the results from our unsupervised method and those from PCA as comparison. The symbols  $N$, $F$, $P$ represent
nematic, ferromagnetic and paramagnetic phases, respectively.
 The dashed lines are data from
the MC simulations~\cite{q2mc} mainly of $L=50$ up to $L=300$.
The color indicates the value of $ch$ index obtained from simulation with  the system size  $L=32$. We now discuss the $F$-$P$, $N$-$F$ and $N$-$P$ transitions as follows.
\renewcommand{\theenumi}{\roman{enumi}}%

(i) The $F$-$P$ phase transition:
Interestingly, the positions of  the peaks  by $ch$  are consistent
with  the dashed line of the phase boundary in
the whole region $\Delta>0.4$.
The index $ch$  performs  very well where  $\Delta$  is away from the pure XY model limit. The essential nature of the $F$-$P$  phase transition is still  BKT.
%For example, when the $\Delta=0$, the model will sit
%at the nematic phase. The phase transition from the nematic phase
%to the paramagnetic phase is different because the vortex is
%5half rather than integer vortex excitation.

%Therefore, $ch$ index is a very good indicator for the phase
%boundary condition at $\Delta=1$, namely the pure XY model.

%Even for the tri-critical regimes $\Delta=0.3$, the position of the peaks just derivate 0.1 from the  known boundary conditions.

%Two possible way of improve the accuracy.
%First one is
%increasing the lattice size up to $L=64$, $\Delta=0$, 0.1 as shown
%in Figs.~\ref{q2ph} (b) and (c).

%In each subfigures,
%we draw four lines, and the peaks for $L=64$ are distributed at $T=0.8$ and $T=0.9$, with and with out %half constraint respectively.

(ii) The $N$-$P$ phase transition:
In the regime $\Delta=0$,  the $ch$ peaks around
 $T=0.7$, which is $0.2$ less than $0.9$. This discrepancy is likely due to the nature of half-vorticity in the $N$ phase.
We can improve the result by limiting the configurations to have the half-winding number $\nu_{x(y)}=1/2$ as the topological constraint in our Monte Carlo simulation.  The half-vortex physics has been discussed in Ref.~\cite{q2mc}.

%The $q=2$ GXY model has half-winding number vortex excitation.
%Therefore, for each row in the square lattices,
%Also in the regimes of \tony{critical} temperature,
%we ingonore the ferromagentic phase, namely,
%the spin orientation are equal between different
%lattice sites.
Thus, we only consider ($\nu_x$,$\nu_y$) in the four types of  combinations ($\pm 1/2$, 0) and (0, $\pm 1/2$).
To form  ($\nu_x=1/2$,0), the difference between a pair  of
spins located at the left most and right most boundaries is fixed as $\pi$ and we assign each spin using
the Eq.~(\ref{eq:1d}).
With the half vortex constraint,  our results illustrated by the red symbols move closer to
the dashed lines  of the $N-P$ transition in Fig.~\ref{q2ph} (a) than
 the results using the integer-value constraint.

Figs.~\ref{q2ph} (b) and (c) show the details at
$\Delta=0$ and $\Delta=0.1$, respectively. For $L=64$, we
generate two groups of data and the peaks almost converge
in the interval $[0.82,0.85]$, and the convergence is closer to $0.89$ 
%\tony{0.89 is outside the interval 0.82..0.85.  What do you mean? Are you saying the convergence is closer to 0.89?} 
than the result from the integer vortex constraint.  For $\Delta=0.1$, the half vortex constraint gives better results
 at $T_c =0.8$.
%
%Without half vortex constraint, we get \tony{$Tc=0.9$ $Tc$ or $T_c$?}  for
%$\Delta$ = 0, 0.1  with $L=64$.

(iii) The $N$-$F$ phase transition:
When we realize that the topological constraint makes
the peaks (features) of the distribution for the spin directions
implicit
in the $F$ and $P$ phases, we use the configurations without any constraint
in this case.  The results are labeled by purple
symbols with legend `$pca$, $L32$'.

\begin{figure}[htb]
\includegraphics[width=0.8\linewidth]{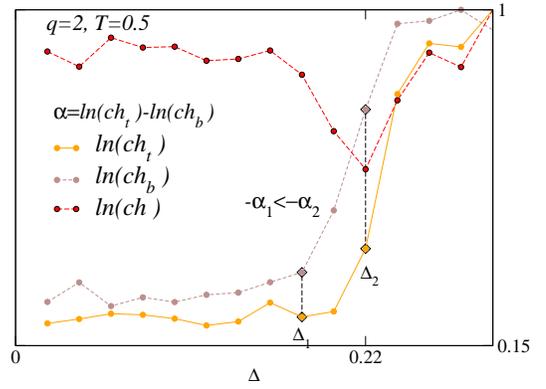}
\caption{For the $q=2$ GXY model, $ch$, $ch_t$, $ch_b$ in log-scale vs. $\Delta$ without any constraint.
$ch$ is obtained by applying k-PCA and k-means. }
%\tony{by perform? What do you mean? by applying k-PCA and k-means?} after k-PCA and k-means.}
 \label{fig:logch}
\end{figure}

The behavior of $ch$ depends on the structure of the sample points, and sometimes $ch$ emerges as a sharp jump at the phase transition point such as  in the Ising model discussed in
the main text. Sometimes it is a local minimum value
at the phase transition point.
Take the  $q=2$ GXY model at $T=0.5$ as an example,
in Fig.~\ref{fig:logch}, both $\ln (ch_t)$ and  $\ln (ch_b)$ increase as a function of $\Delta$. However
the difference
$\alpha= \ln (ch_t)- \ln (ch_b)=-\mbox{gap}$
  decreases  first and then increases around
$\Delta_2=0.22$, where the gap has a positive value. 
%\tony{the gap is the absolute value?? What does that mean?  Do you mean a positive value?}.
\vskip 0.5 cm
According to the following equation:
\begin{equation}
ch= \exp( \ln (ch_t)- \ln (ch_b))=\exp (-\mbox{gap}),
\end{equation}
clearly, $min(ch)\Rightarrow \max(\mbox{gap})$ and
the local minimum of
the $ch$ is at the location of $\Delta_2$.

\begin{figure}[bt]
	\includegraphics[width=0.90\linewidth]{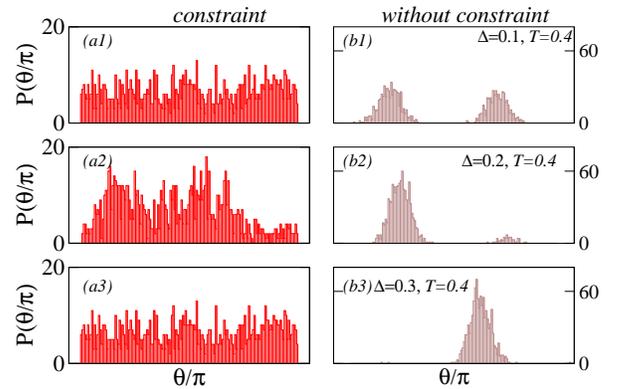}
\caption{In the $N$ ($\Delta=0.1$) and  $F$ ($\Delta=0.3$) phases 
%\tony{In the $N$ ($\Delta=0.1$) phase,  $F$ ($\Delta=0.3$) phase - What do you mean? in both phases? ``In the $N$ ($\Delta=0.1$) and  $F$ ($\Delta=0.3$) phases''?}  
and critical points $\Delta=0.2$,  the distributions of spin
vector of the $q = 2$ GXY model with (a1)-(a3) and without
(b1)-(b3) constraints, respectively. }
\label{fig:3s}
\end{figure}

 For the $N$-$F$ phase transition in Fig.~\ref{q2ph} (a), the PCA is stronger
than the DM method. Here, we would like to  give a physical
discussion about this because  such a difference is related to
the advantage of the proposed method, and thus a more detailed
discussion should be made.

%\noindent \textcolor[rgb]{0.00,0.07,1.00}{A5}:
The reasons for PCA being stronger for the $N-F$ transition
can be explained as follows:
\renewcommand{\theenumi}{\roman{enumi}}%

\begin{figure}[htb]
	\includegraphics[width=0.90\linewidth]{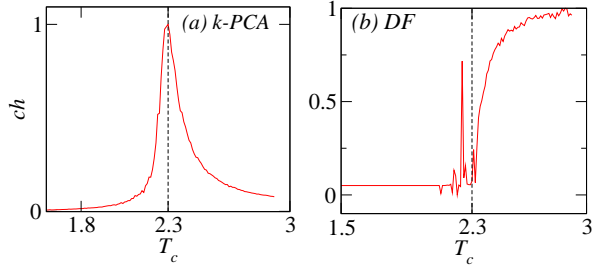}
\caption{
(Color online) The value of $ch_b$ obtained by kernel PCA (k-PCA) and
the DM model for the complete same configurations of the $64\times 64$ Ising model
without any constraint.}
\label{fig:isd}
\end{figure}

(i) The $N-F$ phase transition  is not a topological phase transition\cite{q8mc}.
The DM method designed here is  to determine the topological phase
transition.
It is still interesting to see the $ch$ of DM by inputting Ising configurations without any
topological constraint.
In Fig.~\ref{fig:isd}, using k-PCA and the DM method with complete same configurations,  the signatures of phase transition emerge and
 the results are consistent at $T_c =2.3$.
However, the signature of k-PCA method is clearer.

%For the more complex data (configurations from GXY model) of the $N-F$ phase transition,
%the peaks by  PCA method is already become weak ( not very sharp). It is
%understandable that DF method does not work at all. \\

%Further more,  according to the previous Ref.~\cite{q8mc}, $N-F$ phase transition is a
 %Ising universality and PCA method is enough to obtain the phase transition. \\

(ii) From the view of the data, it is also understandable that
PCA is stronger.
Without any topological constraint, in the nematic ($N$) phase, the spins
prefer two dominant directions and their histograms
obey a double peaks structure. In the Ferromagnetic ($F$)  phase,
one main direction remains and one peak emerges for the histograms.
Fig.~\ref{fig:3s} (b1)-(b3) show the typical histograms in three phases.
The main feature difference between the  phases is obvious.

However, with a topological constraint, the spin directions are mainly distributed according to the winding numbers. For example, the spin angles in each row  are $\theta_{x}=2\pi x/L$
with additional fluctuations,
where $x$ and $L$ are the number index and total number, respectively, in each row.
Therefore, the distribution sits almost in the range $[0,2 \pi ]$ as shown in
Figs.~\ref{fig:3s}(a1)-(a3).
The feature difference of spin angles  disappears  when using the constraint.
Therefore, the DM method cannot get transition points using the configurations with  constraints.  \\

\vskip 0.5 cm

\begin{figure}[htb]
\includegraphics[width=0.95\linewidth]{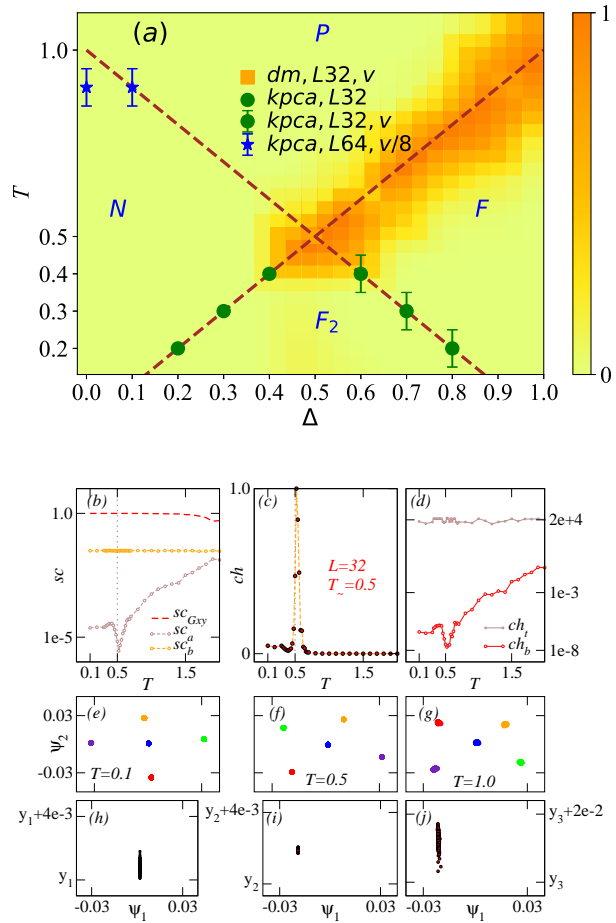}
\vskip 0.5cm
\includegraphics[width=0.9\linewidth]{fig11b-ch_updown.eps}
\vskip 0.2cm
\caption{(Color online) (a) Phase diagram (with $N$, $F$, $F_2$,  $P$ phases) of the $q=8$ GXY model  and results from  calculating the $ch$ index, whose
values are represented by colors. The dashed blue lines are MC results \cite{q8mc}. (b) The curves of $sc$, $sc_a$, $sc_b$ vs. $T$ at  $\Delta=0.5$.
  (c) and (d) The curves of  $ch$, $ch_t$ and $ch_b$ vs. T, at  $\Delta$=0.5 (and note that $T_c =0.5$).
(e)-(g): Scatter plot of $\Psi_1$ and $\Psi_2$ at $T=0.1$, $0.5$  and $1.0$. (h)-(j): Zoom-in view for a fixed group of data.}

\label{q8phchss}
\end{figure}

\subsection{\textbf{\textit{q=8}} GXY model}

%\smallskip {\it {Two dimensional $q=8$ GXY model.--}}
For the $q=8$ GXY model \cite{q8mc}, Fig.~\ref{q8phchss} (a) shows the global phase diagram using the values of the $ch$ index.
The phases $N$, $F_2$, $F$, $P$ are obtained and the distributions are also
shown.

In the new $F_2$ phase, the distribution of the spin vectors has 8 peaks, but is dominated by $4$ possible angles. The `X' shape dashed lines are from Refs.~\cite{q2mc,q8mc}.  The orange color represents the values of $ch$ by the DM method.

Let us first discuss the $F$($F_2$)-$P$ transition.
Clearly,  for $F$-$P$ transition in the  range $0.5<\Delta<1$,  the peak positions of $ch$ align well with the dashed line.
In the center of 'X', the $F_2$ and $P$ transition are also consistent with MC result i.e.,  $T_c =0.5$.

However, we could not use the intersection
of the cluster average distance $\bar{D}$ and within-cluster dispersion $\bar{\sigma}$ as described in Ref.~\cite{nature} because $\bar{D}$  does not  vary too much and there is no intersection as shown in Figs.~\ref{q8phchss} (e)-(g). The five different colors therein represent the five typical topological sectors.
However, fortunately, when  zooming in the figures  in  Figs.~\ref{q8phchss} (g)-(i), we find that, near the transition region,  the shape of a fixed
cluster shrinks   because  the data points gather closer together, and hence the within-cluster dispersion $\bar{\sigma}$ is smaller.
The index $ch =\frac{ch_t}{ch_b}$ thus develops
a peak around $T_c =0.5$  as shown in Fig.~\ref{q8phchss} (c), with
 $ch_t$ and $ch_b$ also displayed in Fig.~\ref{q8phchss} (d).
%{\bc The local minimum peak of $ch_b$ \tony{can} be understand and explained in the appendix.}
In contrast, Fig.~\ref{q8phchss} (b)  shows
that $sc$ cannot be used to signal the transition temperature $T_c$ because it evaluates
the difference, $sc_b-sc_a$, but  $sc_a \ll sc_b$ in spite of the fact
that  $sc_a$ has a local minimum.

It should be mentioned that for the $N$-$P$ transition,
 the use of either the integer or half vortex constraint is not suitable.  Instead the $\nu_{x(y)}= 1/8$ vortex constraint is needed in generating  the configurations. Moreover, after comparison, we find that
kPCA works better if we use {$\theta_i$} as the input into the kPCA (using PCA does not work well).  The kernel  used for kPCA
defined as a radial basis functional kernel $\exp (-\gamma||x/L-y/L||^2)$, where $L=32$ is the system size, $x$ and $y$ are the configurations {$\{\theta_i\}$} and $\gamma=1$ is the default value~\cite{gp}.

The other details of the $F$-$P$,  $F_2$-$F$, $N$-$F_2$ and  $N$-$P$ transitions will be discussed as follows.
%\tony{in order? What do you mean? Do you mean ``as follows''?}.
\begin{figure}[tb]
\includegraphics[width=0.9\linewidth]{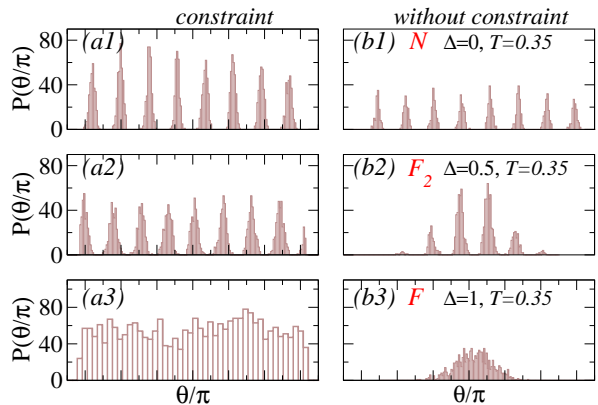}
\caption{In the $N$, $F_2$, $F$ phases, the distributions of spin
vector of the $q=8$ GXY model with (a1)-(a3) and
without (b1)-(b3) constraints, respectively.}
\label{fig:8peak}
\end{figure}
For the $q=8$ GXY model,
Figs.~\ref{fig:8peak} show the distributions
of spin vectors in the
(a1) $N$, (a2) $F_2$ and (a3) $F$ phases with the integer-vortex constraint.
The distributions of spin vectors in the $N$ and $F_2$ phase
have no obvious differences. To distinguish the $N$ and $F_2$ phases, the constraints are canceled  and the distributions are shown in Figs.~\ref{fig:8peak} (b1)-(b3)
respectively.

\begin{figure}[tbp!]
\includegraphics[width=0.9\linewidth]{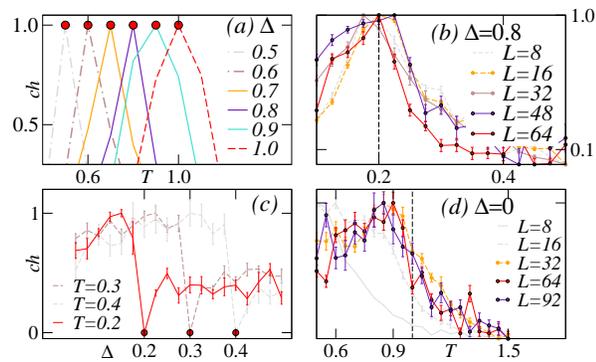}
\caption{ (Color online) (a) $F$-$P$: $ch$ vs $T$ at
 fixed $\Delta$ in the interval $[0.5,1]$. The peaks are at $0.5$, $0.6$, $0.7$, $0.8$, $0.9$  and  1, respectively.
(b) $F_2$-$F$: $ch$ vs. $T$ and  $T_c$ is  $0.2$ for $\Delta=0.8$.  (c) the numerator $ch_t$  and  denominator $ch_b$
of the $ch$ index in Eq.~(\ref{eq:ch}). (c) $N$-$F_2$: $ch$  vs. $\Delta$ for $T=0.2$, $0.3$ and $0.4$.
(d) $N$-$P$: using the $\frac{1}{8}$ vortex constraint and kPCA,  $ch$ vs. $T$ and $T_c$ is about $0.9$.
 }

	\label{fig:q8f}
\end{figure}

Figs.~\ref{fig:q8f} (a)-(d) show the detail of phase transition $F$-$P$, $F_2$-$F$, $N$-$F_2$, respectively. For the $F$-$P$ transition, in Fig.~\ref{fig:q8f} (a), fixing
$\Delta$ in the interval $[0.5,1]$ at steps of $0.1$, the peaks of
$ch$ are located at $0.5$ and $1$ respectively, 
% \tony{$\Delta$=0.5 to $\Delta=1$ with interval $0.1$, the peaks of $ch$ locate at 0.5 to 1 respectively, - Terribly explained! Do you mean ``fixing $\Delta$ in the interval $[0.5,1]$ at steps of $0.1$, the peaks of $ch$ are located at $0.5$ and $1$ respectively'',?} 
completely matching the dashed lines  in the global phase diagram in Fig.\ref{q8phchss}.

In Fig.~\ref{fig:q8f} (b), for the $F_2$-$F$ transition,  by fixing $\Delta=0.8$, the peaks of
$ch$ are located at $0.2$ with $L=8$, $16$, $32$, $48$ and $64$.
The other values of $\Delta$ are not shown.
In Fig.~\ref{fig:q8f} (c), fixing $T$=$0.2$, $0.3$ and $0.4$, the positions $\Delta$ of local minimum of
$ch$ located at $0.2$, $0.3$ and $0.4$ respectively.
% \tony{Note: these values are the same i.e T=ch?}.

In Fig.\ref{fig:q8f} (d), for $\Delta=0$,
using the $1/8$ vortex constraint results in the most accurate critical points.
%\tony{more accurate the critical points - Makes NO sense - do you mean ``the most accurate critical points''?}. 
The peak position is at  $0.9$
 more closely to $1$ than $0.7$ by the integer vortex constraint.
The error bars are
obtained by the bootstrap method using 400 randomly chosen configurations between the total 2000 configurations and total of 20 bins.

%
%Therefore,  the value of
%Sc are shown as a function of $T$ in (b). Considering the
%performace of SC means the phase transition, we mark
%mark the place to begin to decrease from the maximum value 1 with symbols.
%The position are also labeled by the red squares in the phase diagram,
%consistent completely with the dashed line.
%
% In Fig.~\ref{fig:q8f}   (c), we choose  $\Delta=0.5$, $0.6$, $0.7$, $0.8$, $0.9$ and $1$, the peaks of the curve  $ch$ vs. $T$ are located at
%$T_c$ = $0.5$, $0.6$, $0.7$, $0.8$, $0.9$ and 1 respectively \tony{these values are the same i.e $\Delta = Tc$ or $T_c$?},
%completely consistent with the dashed line.

%Since the $ch$ and Sc are index validation of performace of classification to the data.
%Here,  Fig.~\ref{scatter} directly shows the first two components
%$\psi_1$ and $\psi_2$  in the eigenvectors $\Psi$ corresponding to %the matrix $P$.
%The five types of colors labels the five classes by the K-Mean method or
%DBSCAN method. {\bc We will show the improvement of DBScan to the %k-means in
%the appendix.}

% The temperatures
%are $T= 0.9$ , 1.0 and 1.1, respectively for the each column. The fisrt row is for $\Delta=0$,
%and the second row shows the 60 times  zoomed ( magnified figure ) in $\psi_2$ direction.
%In the critical temperature $T_c =1.0$, the within-cluster dispersing is smallest, which is corresponding
%to the shortest distribution in the shape.

%The advantage of $ch$ index over the SS????????

%In Fig

\section{Other technical modifications}
%{\color{red} ??there should be a brief summary on conclusion here.}
In the above sections, during the use of the k-means method, we  apply two output
eigenvectors of the diffusion matrix $P$ and then get the values of the $ch$
index. The conclusion is  that using two leading vectors leads
to the best accuracy.  At the same time, it is possible to design
a way to  determine the super parameter $\varepsilon/\varepsilon_0$ automatically.
In this section, we will focus on such issues for the completeness of our method.

\label{sec:otm}
\vskip 0.5 cm
\begin{figure}[htb]
	\includegraphics[width=0.90\linewidth]{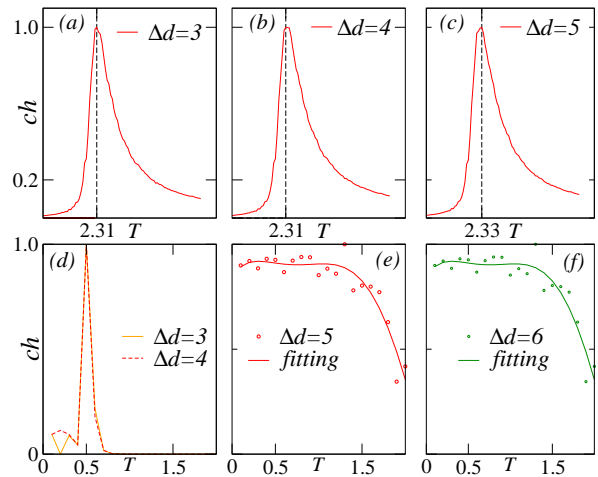}
\caption{ $ch$ index as a function of $T$ for various value of  $\Delta d$  (a) $\Delta d=3$  (b) $\Delta d=4$ (c) $\Delta d=5$ for Ising systems.  $ch$. vs. $T$ of the GXY model at fixed $\Delta =0.5$ and varying $T$ with number of vectors considered (d) $\Delta d=3$ and $4$ (e) $\Delta d=5$ (f) $\Delta d=6$, respectively. The results are best using two vectors $\Delta d=2$. }
\label{fig:hd}
\end{figure}

%\textcolor[rgb]{0.00,0.07,1.00}{A2:\\}
% \noindent \textcolor[rgb]{1.00,0.00,0.00}{Q2}: Why \tony{ars the reduced dimensions} fixed to \tony{two dimensions}? If a higher
%dimensional feature is used when the k-means method is performed, does
%the accuracy change? I think this investigation is useful information
%for readers.\\
Here,  the effects of higher-dimensional features will be discussed using k-means, namely,
whether or not the accuracy is enhanced when including more features
will be clarified.
The answer is that retaining more dimensions of the eigenvectors,
the result may deviate from the right critical points or even predict wrong critical points.

%In the step of k-means, usually just two vectors are considered as input
%to

%\noindent \textcolor[rgb]{0.00,0.07,1.00}{A2}: Thanks to the reviewers suggestions, we have consider higher dimension
% data to calculate the $ch$ score.
Assuming the covariance matrix of the PCA method is $C$,
and its  eigenvectors are $\{\Psi_d\}$,
where $d=1, \cdots , d_{max} (\Delta d=d_{max}) $.
For the Ising model, in Sec.~\ref{sec:ising}, the
first and second vectors $\{\Psi_d\}$, are used,
where $ d_{max}=2$, or  $\Delta d=2$.
Here, we consider vectors with $\Delta d= 3$, $4$ and $5$.
The difference is found between the results of two-dimensional vectors
and higher dimensional data, as shown in Figs.~\ref{fig:hd}(a)-(c).
The estimated position of peaks are $2.31$, $2.31$ and  $2.33$ respectively. The result $T_c=2.3$ for $\Delta d=2$
is the closest to the peak of the specific heat.

For the generalized XY model ($q=8$, $\Delta=0.5$), with the DM method,   the vectors with index  $d=2,\cdots, d_{max}$  ($\Delta d= d_{max}-1$) are  used
as the first vector ignored.
In Fig.\ref{fig:hd} (d),  with $\Delta d=3$ or $4$,
the position of peaks  are $T_c =0.5$ as the same as that of
$\Delta d=2$. However, if higher-dimensional data are used, such as $\Delta d=5$ and $6$,
the $ch$ score does not yield the correct signature of the phase transition
as shown in Figs.\ref{fig:hd} (e) and (f). \\

Another issue discussed here is whether one can design an automatic method to determine the hyperparameter  $\varepsilon/\varepsilon_0$.
Since its value has been tried many times for some reasonable choices,
one may expect to find a way to determine its optimal value automatically.

%\vskip 0.5 cm
%\begin{figure}[htb]
%	%\includegraphics[width=0.90\linewidth]{getEpsilon.eps}
%\caption{(a)Histogram of $T_c$ obtained by using various $\varepsilon_0/\varepsilon$.
%(b) The phase transition points $T_c$ obtained as a function of $\varepsilon_0/\varepsilon$
%from the peaks or second peak through the line $ch$ vs. $T$. }
%\label{fig:3s}
%\end{figure}

To the best of our knowledge, at present, there are no such automatic determination
of the hyperparameter $\varepsilon/\varepsilon_0$.
Here we devise a simple  analysis from the statistics (histogram)
of $T_c$ obtained by various  $\varepsilon/\varepsilon_0$:
\renewcommand{\theenumi}{\roman{enumi}}%
\begin{enumerate}
\item Calculate $ch(T)$ as a function of temperature $T$ with  various $\varepsilon/\varepsilon_0$ in a range from $ \varepsilon^{min}/\varepsilon_0$ to $ \varepsilon^{max}/\varepsilon_0$  by using
the configurations with topological constraints;
\item Store the position of $ch$ peaks, i.e., $T_c$ and the times they appear as shown in Fig.~\ref{fig:chi} (e) and (f).
Sometimes, the peaks may not be not obvious, and a plateau emerges.
The two ends of the plateau will be stored.
\end{enumerate}
The histogram of $T_c$ vs. $ \varepsilon/\varepsilon_0$ can be used to give the best estimate for the transition. Therefore, corresponding to the highest position $T_c=0.96$, the values $4\leq \varepsilon/\varepsilon_0\leq6$
 are acceptable. \\

Another possible approach is to calculate a location dependent $\sigma$ for each data point instead of selecting a single scaling parameter \cite{localsigma}.
Then, the kernel matrix  between a pair of points can be written as
\begin{equation}
\omega(x_{j},x_{j})=\exp (-\frac{||{x_{i}-x_{j}}||^{2}}{\sigma_i\sigma_j}).
\label{eq:sigma}
\end{equation}
where $\sigma_i$ and $\sigma_j$ are the local scale parameters for $x_i$ and $x_j$, respectively. The selection of the local scale $\sigma_i$ is determined by the local statistics of the neighborhood of point $x_i$. For example, the scale can be set as
\begin{equation}
\sigma_i=||{x_{i}-x_{K}}||.
\end{equation}
where $x_K$ is the $K$-th nearest neighbor. However, here, K is also a hyperparameter
to be chosen.
By comparing Eq.~(\ref{eq:sigma}) and Eq.~(\ref{eq:k}),
$2N\varepsilon=\sigma_i \sigma_j$, the obtained $\varepsilon/\varepsilon_0\approx 30 $
is about several times larger than the acceptable regimes.

\section{Discussion and Conclusion}
\label{sec:con}

In summary, we use the Calinski-Harabaz ($ch$) index to successfully locate phase transitions of a few classical statistical models, including the Ising, XY and the generalized XY models. From the scatter data, for which we use the $ch$ index
to obtain the phase transitions for
the Ising, XY and GXY models on lattices. This combines the advantages of the PCA and DM methods, as the scattering data can be based on either the first two leading eigenvectors of the PCA Kernel-PCA or the second and third vectors from DM method.

The advantages of using the $ch$ index are
less steps, wider applications and better convergence.
After k-means applied to the eigenvectors of the diffusion
matrix $P$, using the $ch$ index, we
do not need to maximize the visibility of cluster as proposed in RNS method.
For some phase transitions, it may
not be easy to find the intersections of $\bar {D}$ and $\sigma$.
In Figs.~\ref{q8phchss} (e)-(g), $ch$ index could capture
the signatures of both quantities.
In Fig.~\ref{fig:6res}, for the pure XY
model on the honeycomb lattices, the exact critical point is at $T_c =0.707$.
Using $ch$, when $\varepsilon/\varepsilon_0$ is in the interval [2,6], 
%\tony{2-6  - do you mean in the interval $[2,6]$? then write it that way!}
the estimated results of $T_c$ are all close to $0.7$.

In the pure XY limit, we have tested that  $ch$ can locate the phase transition for the XY model on both the square and honeycomb lattices, similar to the DM method of Rodriguez-Nieva and  Scheurer.
For the $q=2$ GXY model, the values of the $ch$ index in the whole phase
diagram  matched the boundary between the ferromagnetic phase to
paramagnetic phase very well even  away from the pure
XY  limit.  Close to the  $N$-$P$ transition, the accuracy can be improved by  using the $1/2$-vortex constraint in generating the Monte Carlo configurations.
For the $q=8$ GXY model, using intersections of $\bar{D}$ and $\bar{\sigma}$
cannot be used to locate the transition point, such as at $\Delta=0.5$, but
the  $ch$ index can still work to locate the phase transition point due to
its incorporation of  the fluctuation  of samples  within each cluster. Moreover,  the $N$-$P$ transition can also be identified by using the $1/8$ vortex constraint.

We have also compared the results from other indices (see Table~\ref{tab:}), but we find that the $ch$ index works best overall for the models considered in this work. In the future, it will be desirable to systematically study the utility of various parameters in models of statistical physics that exhibit different natures of transitions.

%\noindent\textcolor[rgb]{0.00,0.07,1.00}{ A6}:
The development of applying machine learning to physics could
develop new methods for studying unknown phases.
For the Ising model or similar models, the transition point is well known.
This situation will help to check our proposed method.

If we get the first-hand data from experiments, and do not know the details of
the Hamiltonian, our unsupervised method can help deduce the number of possible classes (phases)
in the data. Furthermore,  using the $ch$ index we could identify phase transition points.
Therefore, the method of using the $ch$ index  is very useful for future non-topological phase transitions.
For the XY model and Hamiltonians similar to the GXY model, by using
the $ch$ index, topological phase transition points were obtained without using the traditional method of
measuring   correlations \cite{q8mc} and  spin stiffness \cite{kao}.
Of course, some reasonable prior knowledge is needed such as the possible winding numbers.
The idea of DM can be applied to topological quantum
systems (see Refs.~\cite{band, dff}). In principle, the topological quantum phases and transitions between them
may be probed using the $ch$ or $sc$ indices.
%For the parameter regimes where $ch$ is not so good, the Silhouette index play very well.
\begin{acknowledgments}
We thank for the valuable discussion with J. F. Rodriguez-Nieva, M. S. Scheurer,
Y. Z. You and T. Chotibut on simulations. We also thank Tony C Scott for his
help.
% TCS: please don't mention proof-reading - that's bad for all of us :-)
TCW is supported by the
National Science Foundation under Grant No. PHY 1915165 J. Wang and H. Tian are supported by NSFC under Grant No. 51901152.
W. Zhang is supported from  Science Foundation for Youths of Shanxi Province under Grant No. 201901D211082;
\end{acknowledgments}

%\end{CJK}
\appendix

\section{The 1D XY model}
\label{sec:1dxy}
The Hamiltonian of the pure XY model reads
\begin{equation}
\label{eqh}
	H=-J\,\sum_{i=1}^{N}\cos(\theta_i-\theta_{i+1})
\end{equation}
where $J$ is the coupling  strength of the nearest neighboring pair  of spins $\langle i, i+1 \rangle$ and throughout this work, we set $J=1$ for simplicity and use the periodic boundary
condition $\theta_{N+1}=\theta$.
For simplicity, following Ref.~\cite{nature}, we generate the spin vectors according to the following,
\begin{equation}
\theta ^{(l)}_{i}=2\pi \nu ^{(l)}i/N+\delta \theta ^
{(l)}_{i} + \overline{\theta}^{(l)},
\label{eq:1d}
\end{equation}
where $\nu$ is the winding number, $l$ is the identification number of the configuration \{$\theta _i$\}, with $i$ varying from 1 to the total length $N$.
The first term $2\pi \nu ^{(l)}i/N$ is used to define 
%\tony{help? What do you mean ``help''?}  
the winding number
$\nu=\sum_{i} \Delta_i/2\pi$, where the $\Delta_i$ is in the range $[-\pi,\pi)$ by the so-called saw  function~\cite{rogerxy} obtained by replacing
$\Delta_i$ with $\Delta_i\pm 2\pi$ if it is not in the target range.
The term $\delta \theta ^{(l)}_{i}$ obeys the Gaussian fluctuation and
$\overline{\theta}^{(l)}$ is generated randomly between [0, $2\pi$).

\begin{figure}[hbt]
\includegraphics[width=0.8\linewidth, height=0.7\linewidth]{fig15a_1d.eps}
\includegraphics[width=0.99\linewidth]{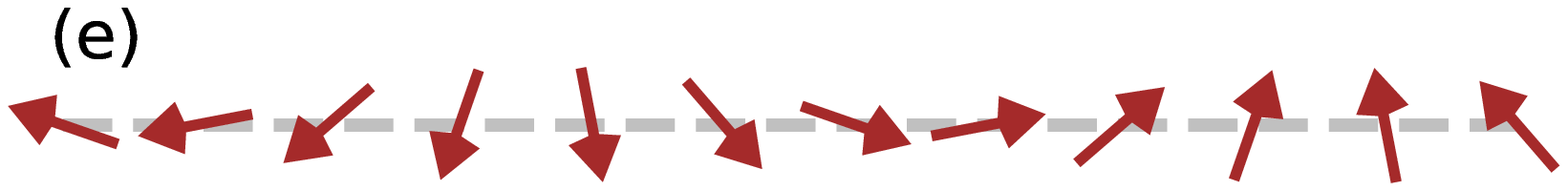}
\caption{(Color online) (a) Histogram count of $\psi_1\sqrt{m}$. (b) The largest 20 eigenvalues of the matrix $P$
(c) $\nu$ vs. $\psi_1$ for the configurations with seven topological sectors (d) The largest $30$ eigenvalues of the matrix $P$.
Clearly in (b) and (d), the leading $2$ and $7$ eigenvalues occur respectively. }
\label{1d}
\end{figure}

We consider two types of generated configurations with winding numbers $\nu=0$ and $\nu=1$.
The histogram of the values of the first component of the diffusion map $\psi_1$ is shown in Fig.~\ref{1d} (a). The histogram of $\psi_1$ has values  at $\pm 1$
 and $\psi$ is a vector with size $m\times 1$, therefore
  the values are equal to $\pm 1$ when $\psi_1$ is re-scaled by $\sqrt{m}$.
Fig.~\ref{1d}  (b) shows the largest 20 eigenvalues of the transition probability matrix $P_{l,l^{'}}$ in Eq.~(\ref{eq:p}). Two maximum eigenvalues are found equal to unity.
Following Ref.~\cite{nature}, we also test $\nu=7$ according to the following equation,
\begin{equation}
\label{eqt+}
\theta ^{(l)}_{i}=2\pi \nu ^{(l)}i/N+\delta \theta ^
{(l)}_{i} + \overline{\theta}^{(l)} + \eta^{(l)}[1-\cos(2\pi i/N)]
\end{equation}
where $\nu=\left\{ 0,\pm1,\pm2,\pm3 \right \} $. %The equation  show the largest eigenvalues of \tony{the} transition probability matrix $P_{l,l^{'}}$ in
%Eq.~(\ref{eqt+}).
%Clearly, we could observed first seven \tony{eigenvalues} are equal 1 in Fig.~\ref{1d} (d) \textbf{\textcolor[rgb]{1.00,0.00,0.00}{and the distribution of Fig}.}~\ref{1d} (c).
%The results are shown in Fig.~\ref{1d} (c) and (d).
We find  that $\psi_1$ has seven district values ranging
from $-0.03$ to $ 0.03$ , corresponding to seven winding numbers marked  by
the symbols in  Fig.~\ref{1d} (c).  Eigenvalues $\lambda_k$ vs. $k$ is also shown in  Fig.~\ref{1d} (d). Clearly, the plateau of eigenvalues appears in $k\le7$.

\begin{figure}[htb]
\includegraphics[width=0.99\linewidth]{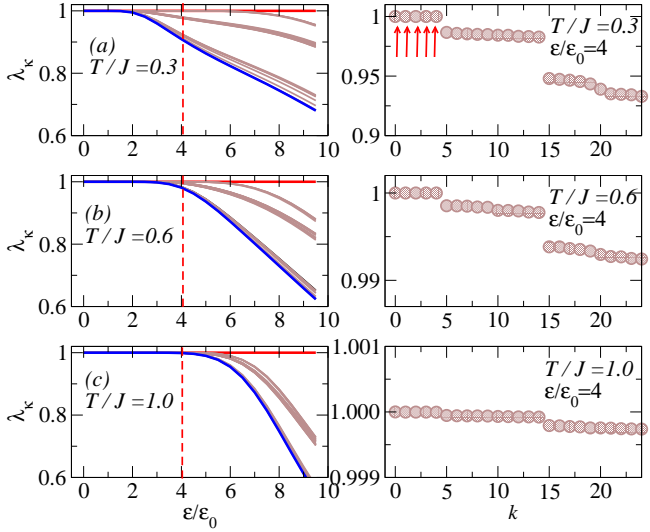}
\caption{(Color online) The distributions of $\lambda_{k}$ vs  $\varepsilon/\varepsilon_{0}$ of the 2D XY model for three typical temperatures (a) $T$=0.3, (b) $T$=0.6 and (c) $T$=1.0. Right column: The corresponding $\lambda_k$ vs $k$. The five arrows point to the eigenvalues of the
leading five topological sectors.}
\label{fig:det}
\end{figure}

%\section{ S3: Local minmum or maximum  of $ch_b$ or $ch_t$}

Figs.~ \ref{fig:det} (a)-(c) show the largest 24  eigenvalues $\lambda_{k\leq 24}$ of $P_{l,l_{'}} $ as a function of $\varepsilon/\varepsilon_0$ for $T=0.3$, 0.6 and
1.0, respectively.
The  band of eigenvalues $\lambda_k$ could not be distinguished for small values of $\varepsilon/\varepsilon_{0}$ and the reason can be seen from the matrix of $P$, which is a diagonal matrix in that limit.
Increasing $\varepsilon/\varepsilon_{0}$,
the band of $\lambda_k$  will separate away from each other.

The  choice of $\varepsilon/\varepsilon_{0}$ is important.
%We try to use the intersections points of $\sigma_{\lambda}$  and $\Delta\lambda$ versus temperature as described  in the preprint version\cite{prep}.
We choose   $\varepsilon/\varepsilon_{0}=4$ marked by
the dashed lines in the left column.  The right column presents
$\lambda_k$ versus $k$($k=0,1,\cdots$).  Clearly, the gap between the $k=4$ eigenvalue and $k=5$ eigenvalue
%\tony{fifth eigenvalue $\lambda_{n-1}$ - Do you mean the 5-th eigenvalue? if it's the 5-th eigenvalue, should not that be $\lambda_5$? What do you mean?} 
 becomes smaller when increasing temperature and subsequently disappears when $T > T_c$.
%The Intersection means that the phase transition point
%Here we exploit the failure of the algorithm to identify the n topological sectors, which leads to a transition temperature of $T_c/J =1.0 \pm 0.05$.

\section{The PCA and K-PCA}
\label{sec:pca}
\begin{figure}[htb]
\includegraphics[width=0.3\linewidth]{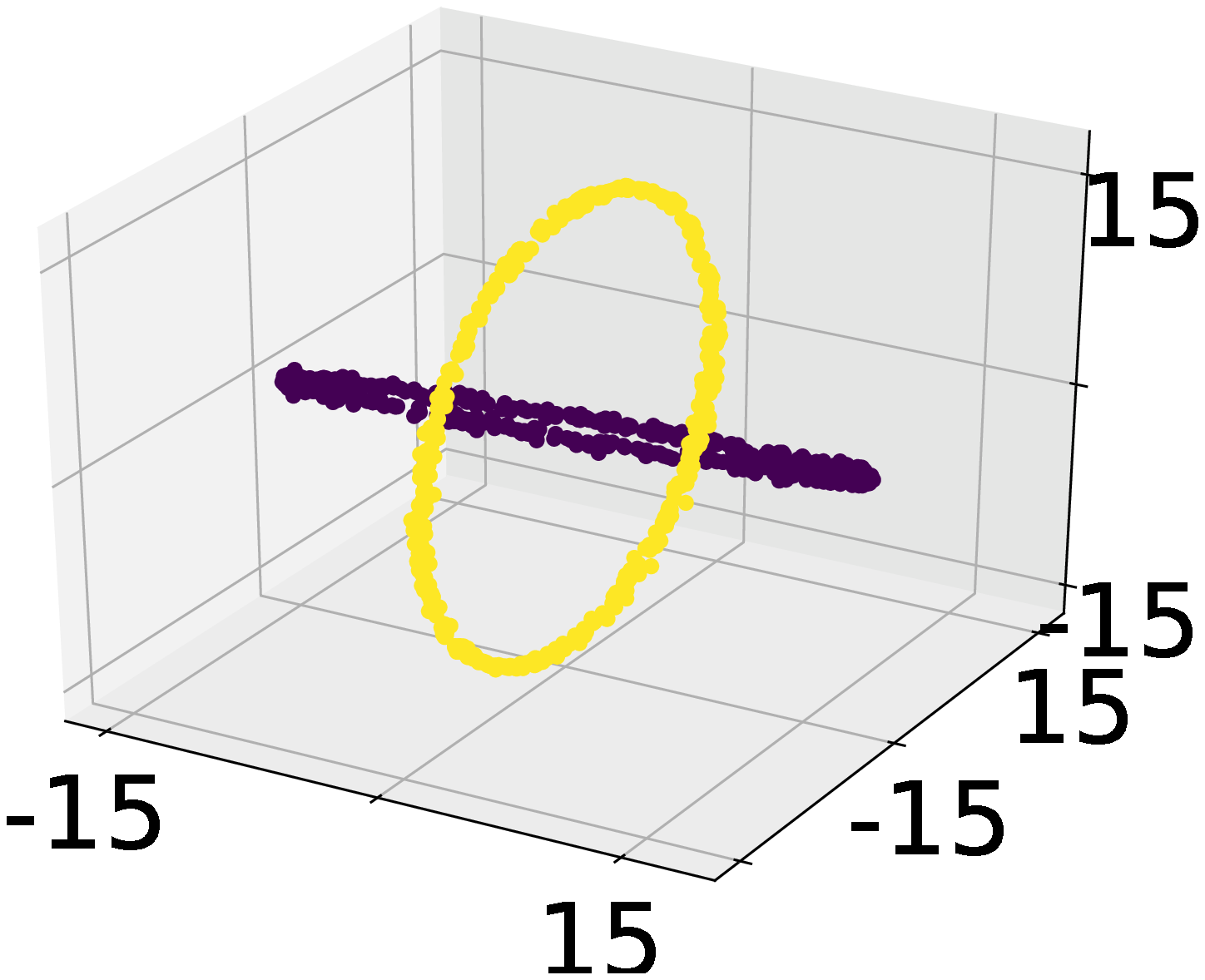}
\includegraphics[width=0.3\linewidth]{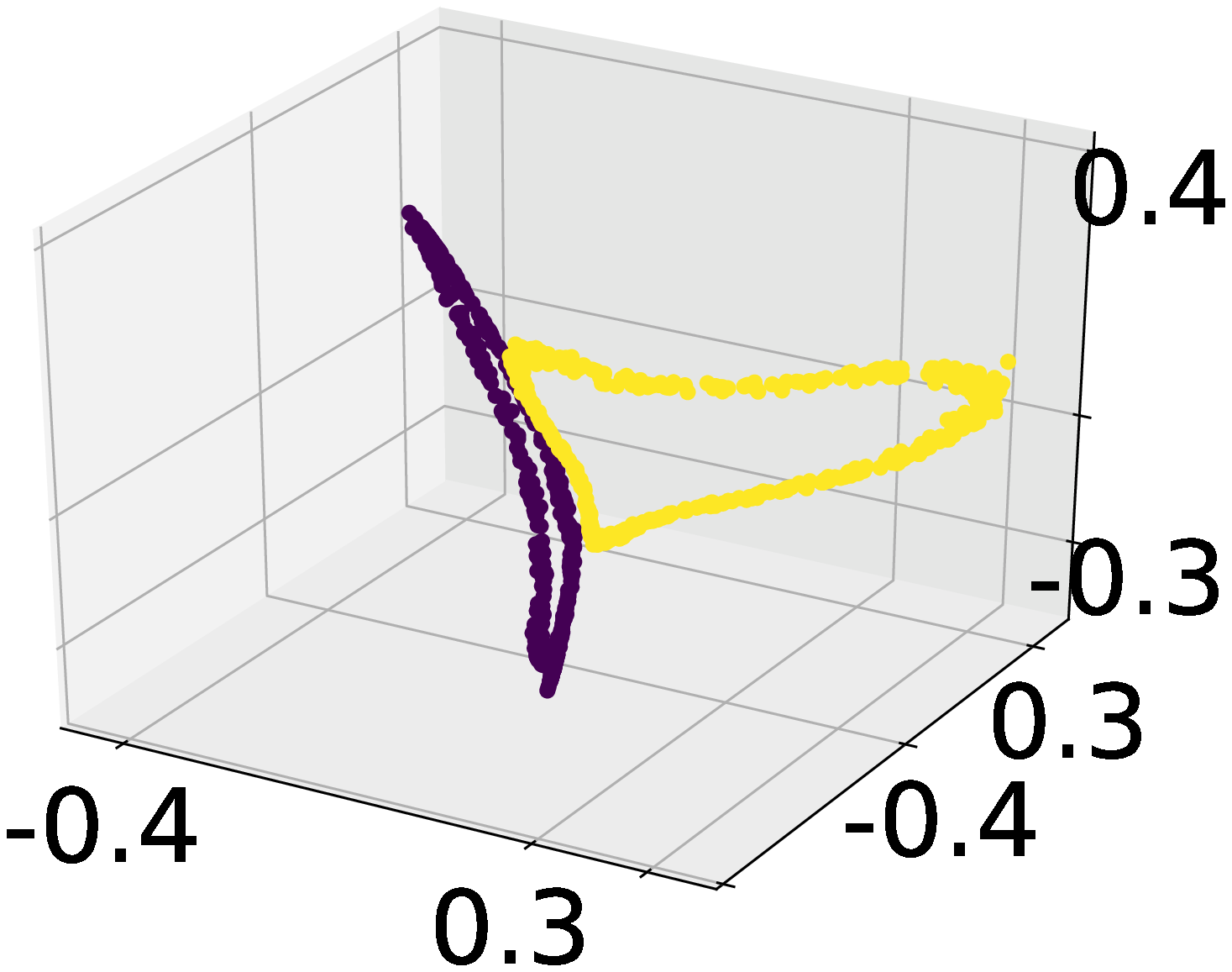}
\includegraphics[width=0.3\linewidth]{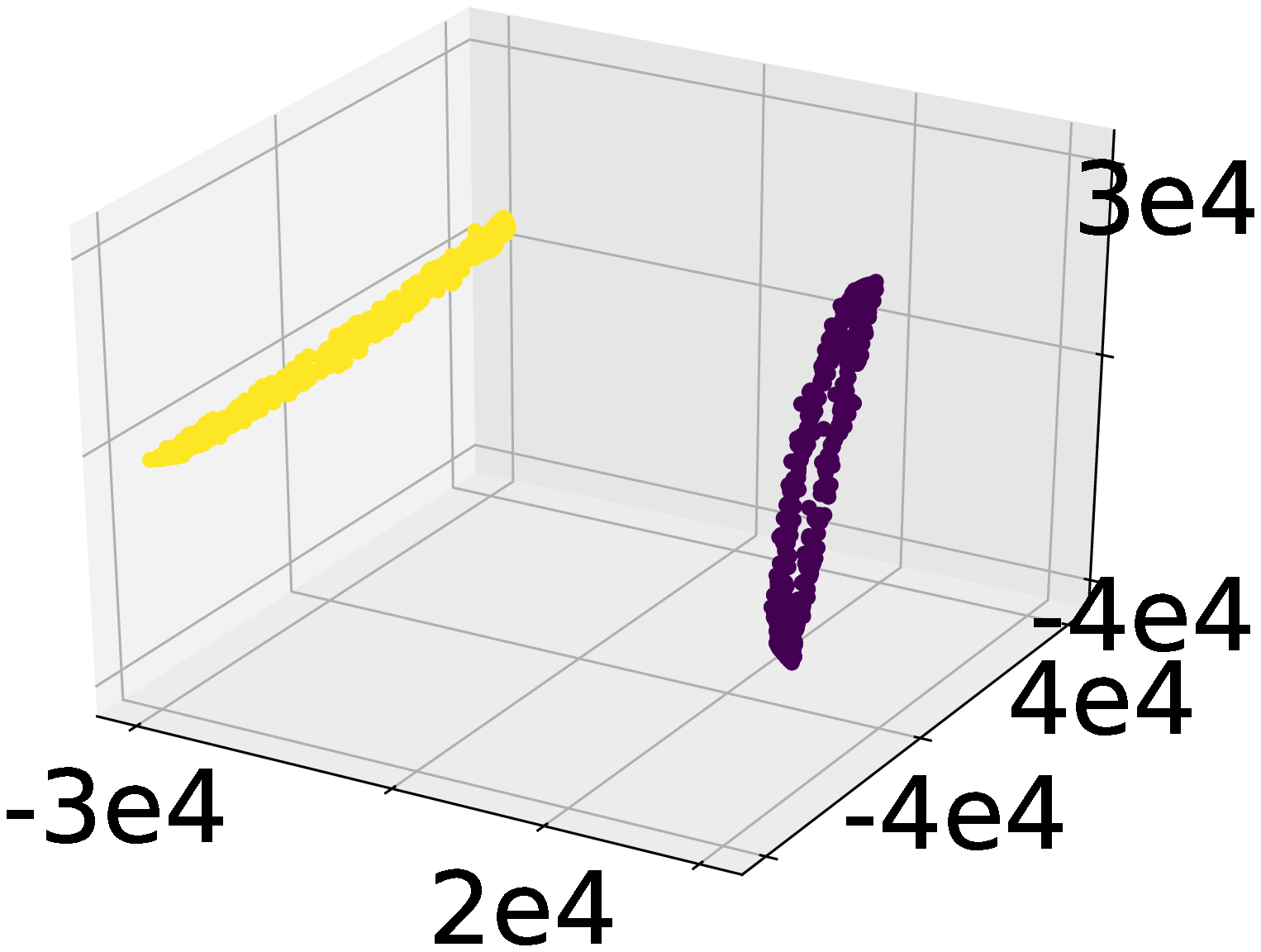}
\includegraphics[width=0.3\linewidth]{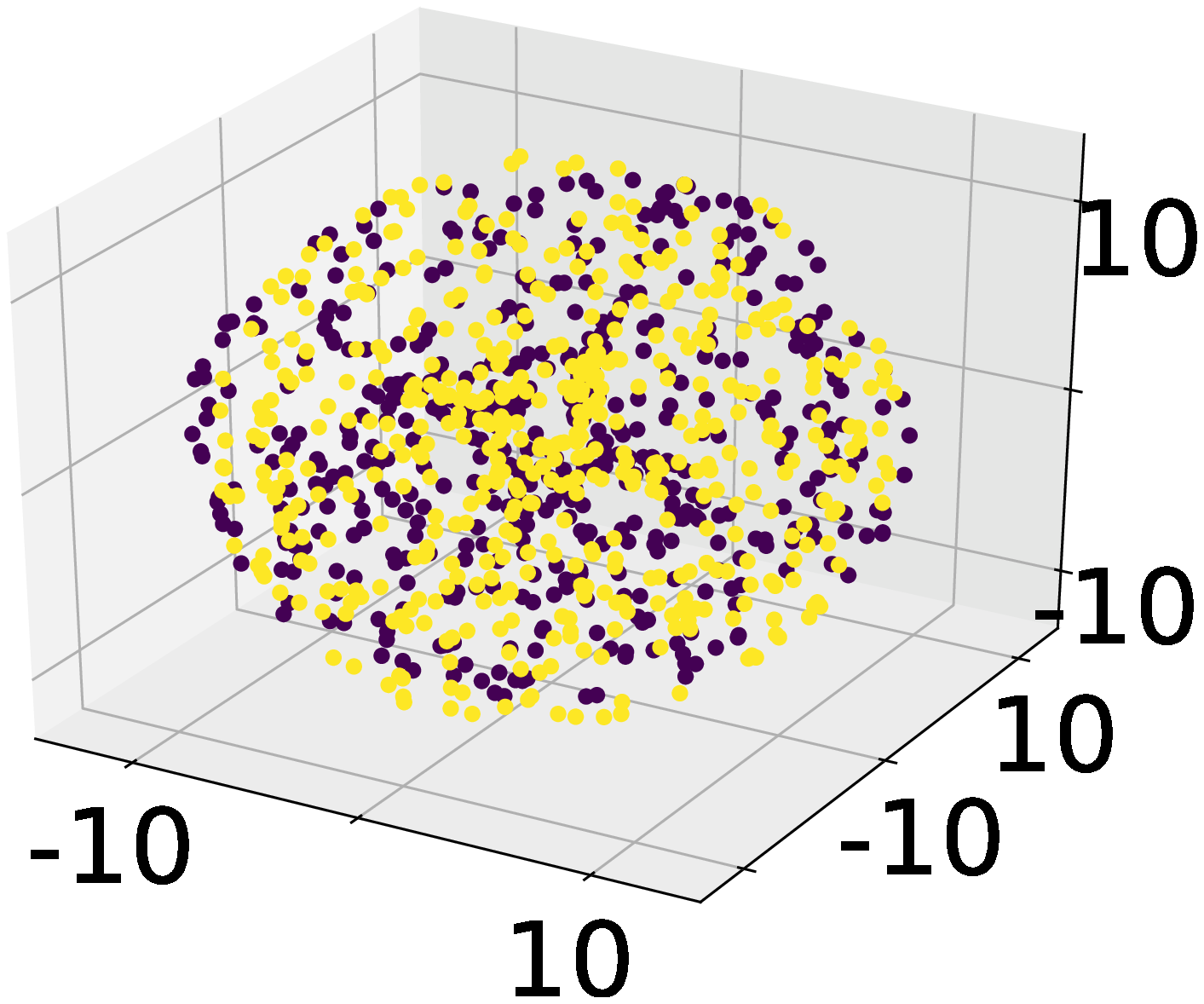}
\includegraphics[width=0.3\linewidth]{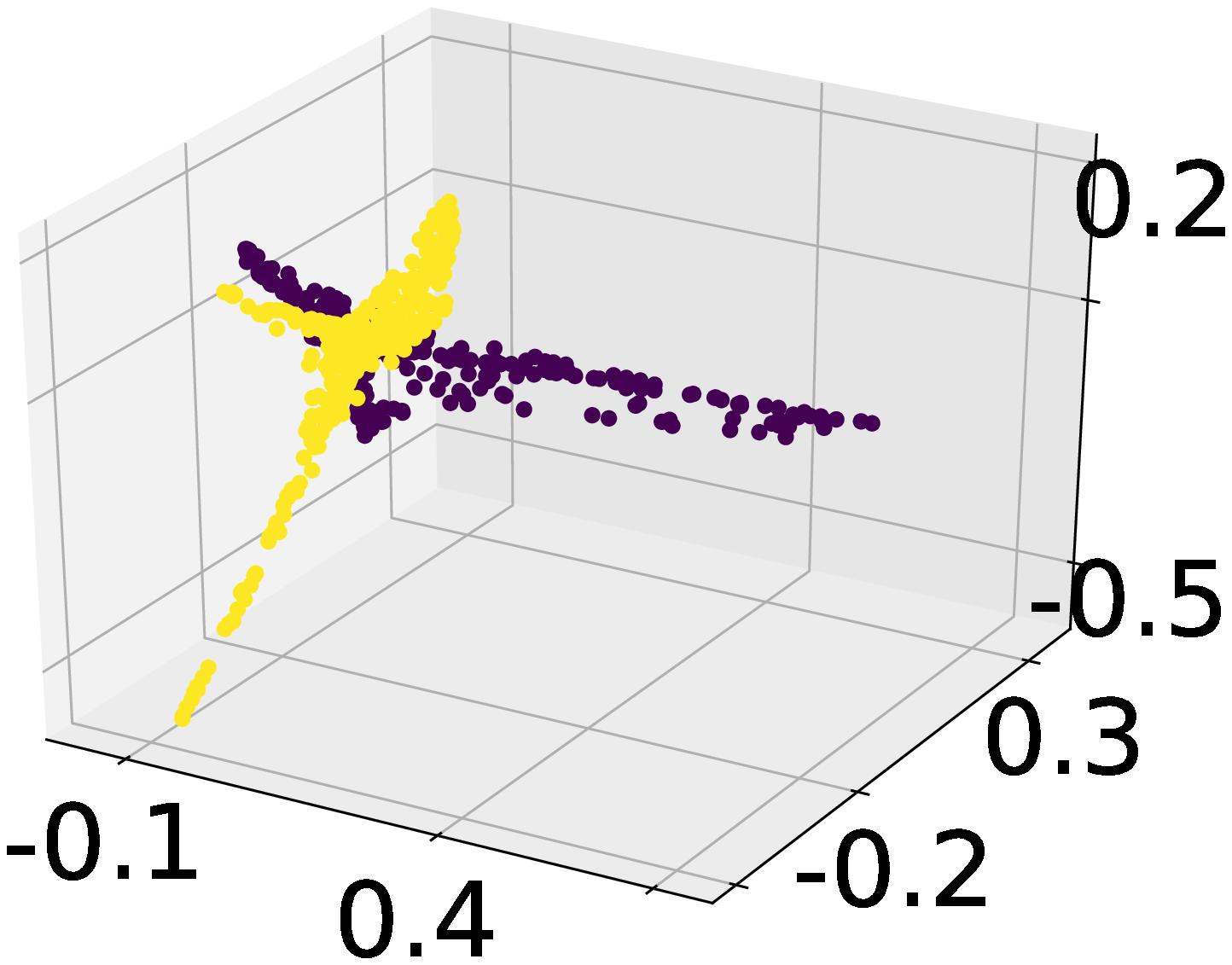}
\includegraphics[width=0.3\linewidth]{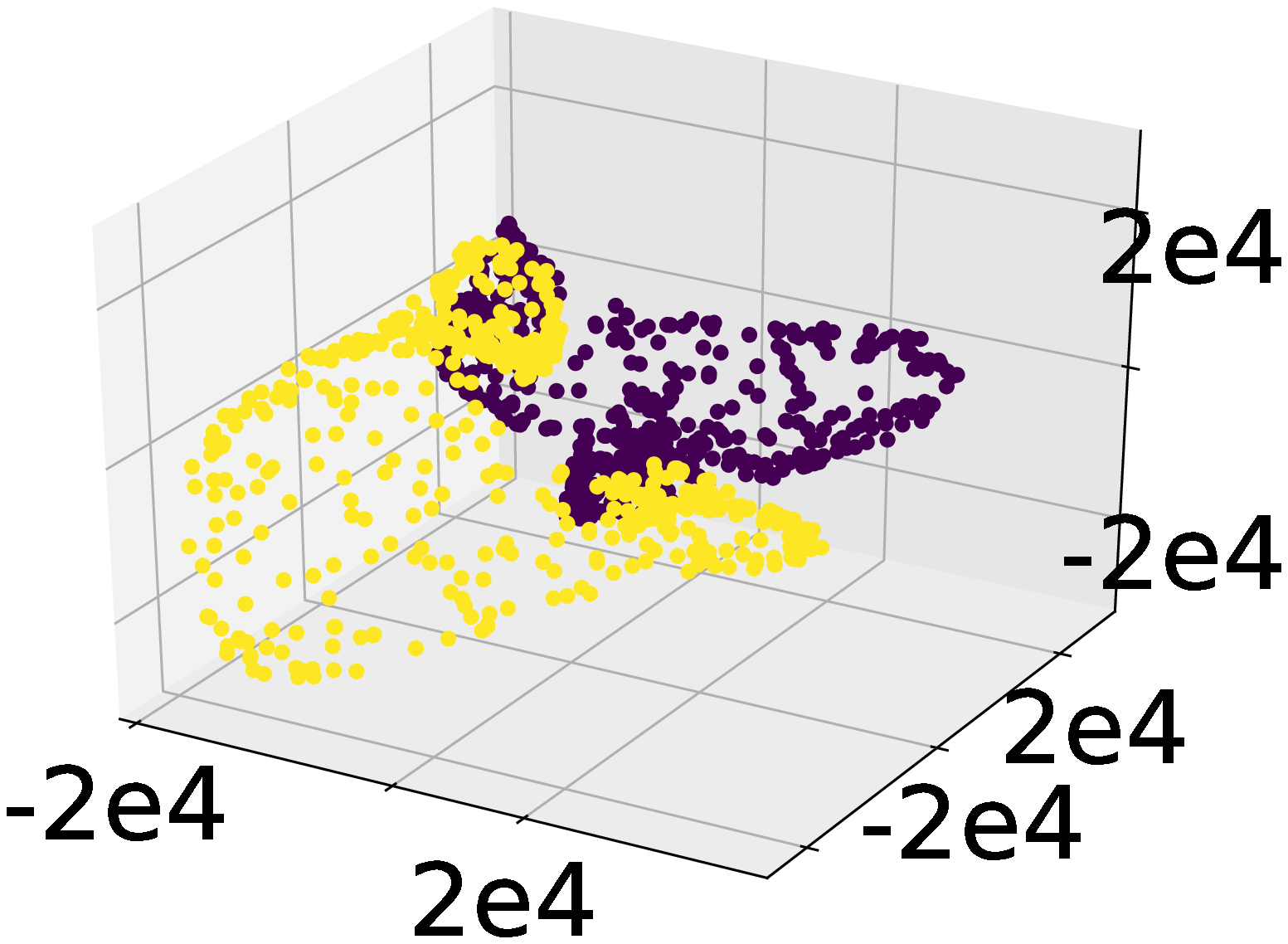}
\caption{(Color online) PCA and kernel PCA with Gaussian and polynomial
kernels  on a simple dataset.
 PCA and kernel PCA with Gaussian and polynomial
kernels  for the generalized dataset. }
 \label{p_kpca}
\end{figure}
Figs.~\ref{p_kpca}, represent some results of PCA and kPCA on
simple and generalized datasets in the appendix of
Ref.~\cite{nature}.

The simple datasets are configurations of one-dimensional XY model with size $N = 256$, $m = 1000$, and $\sigma_{\theta} = 0.3$ and with topological winding numbers $\nu=0$ and $\nu=1$ according to Eq.~(\ref{eq:1d}).
 Figure~\ref{p_kpca}  show results of linear PCA, PCA with Gaussian kernels, and PCA with polynomial
kernels, respectively (the top left, top middle and top right panels). Obviously, the classifications of PCA  with a
nonlinear kernel are much more clear for the XY models.
 The three dimensional visualization is based on the three
  reduced components.
However, the above method fails for data generated
with slight modification of Eq.~(\ref{eq:1d}) with an additional term $ \eta^{(l)}[1-cos(2\pi i/N )] $, where  $\eta^{(l)}$ is random in the range [$-\eta_0$, $\eta_0$]. The results are shown in   Fig.~\ref{p_kpca} at the bottom left, bottom middle and bottom right panels,
respectively.

\section{Other indices}
\label{sec:11ind}
\begin{figure}[hbt]
\includegraphics[width=0.99\linewidth]{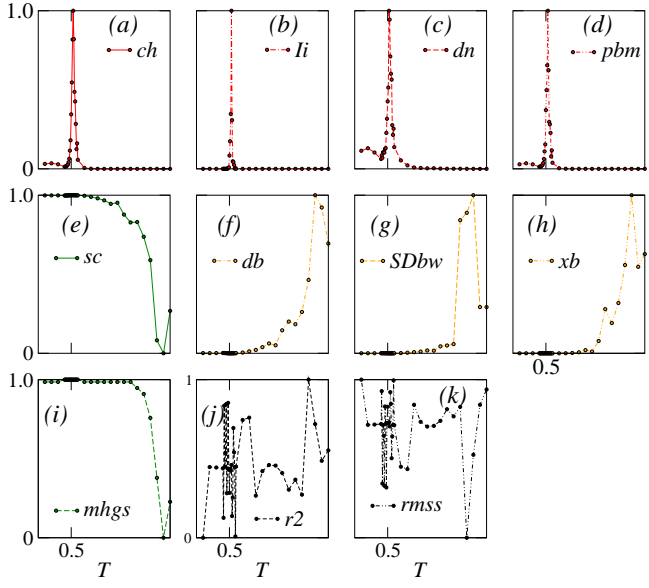}
\caption{In the first row (a)-(d), the indices, called
$ch$, $Ii$, $dn$, $pbm$, can distinguish the $F_2$ and $P$ phases.
(e)-(k) The remaining indices fail to present signals. }
\label{fig:11}
\end{figure}

As listed in Ref.~\cite{11index}, we check the 11 indices listed in Table (\ref{tab})  for validating the classifications. Take the $q=8$ GXY model as an example, the value of the  parameters such as the temperature $T$ and $\Delta$ are fixed  as  those in Fig.~\ref{q8phchss} (b). We find only four indices, presented in Fig.\ref{fig:11} (a)-(d)
produce signals at the critical points. These are $ch$, $Ii$, $dn$ and $pbm$, whose
full names are listed in  Table (\ref{tab}).

The other indices could not
give correct signals to locate the phase transition points.

\begin{table*}
  \caption{INTERNAL CLUSTERING VALIDATION MEASURES.
  }
  \label{tab:}
    \begin{tabular}{clllll}\toprule
         & measure& Notation& Definiton&  \\ \midrule

       1 & Calinski-Harabasz index &  $ch$&
          $\frac{\sum_i n_i d^2 (c_i, c)/(NC-1) }{\sum_i \sum_{x \in C_i} d^2 (x,c_i)/(n-NC) }$
           &\\  \midrule

       2 & Silhouette index&         $sc$&
           \tabincell{l}{
           $
           \left \{

           \frac{1}{NC}\sum_i[\frac{1}{n_i}\sum_{x\in C_i}\frac{b(i)-a(i)}{max(b(i),a(i))}]
           \right \}
           $
           \\
            $
           a(x)=\frac{1}{n_i-1}\sum_{y\in C_i,y\neq x}d(x,y)$\\
           $b(x)=min_{j.j\neq i}[\frac{1}{n_j}\sum_{y\in C_j}d(x,y)]

           $
           }
           & \\ \midrule

       3 & Davies-Bouldin index &     $db$&
           $
\left \{
%\frac{1}{N}  { \sum_{i}max_{j,j\neq i}  [ \frac{1}{n_i} \sum_{x\in C_i}d(x,c_i) + \frac{1}{n_j}d(x,c_j) ]} / d(c_i,c_j)
\frac{1}{NC}    \sum_{i}max_{j,j\neq i}  \frac{  \frac{1}{n_i} \sum_{x\in C_i}d(x,c_i) + \frac{1}{n_j}d(x,c_j) }  { d(c_i,c_j)}
\right \}
           $
           & \\ \midrule

       4 & SDbw validity index &     $SDbw$&
           \tabincell{l}{
           $
           \left \{
           Scat(NC)+Dens\_bw(NC)
           \right \}
           $
           \\
           $
           scat(NC)=\frac{1}{NC}\frac{\sum_i \|\sigma(C_i) \|}{\|\sigma(D)\|}$\\
           $Dens\_bw(N)=$\\
           $\frac{1}{NC(NC-1)}\sum_i[\sum_{j,j\neq i}\frac{\sum_{x\in (C_i\cup C_j)}f(x,u_{ij})}{max\{\sum_{x\in C_i}f(x,c_i),\sum_{x\in C_j}f(x,c_j)\}}]
           $
           }
           &\\ \midrule

       5 & Xie-Beni index &     $xb$ &
           $
           \left \{
           \frac{ \sum_i \sum_{x \in C_i}d^2(x,c_i)} {n\cdot min_{i,j\neq i}d^2(c_i,c_j)}
           \right \}
           $
           &\\ \midrule

       6 & Dunn's indices &     $dn$ &
           $
           \left \{
           min_i[min_j(\frac{min_{x\in C_i,y\in C_j}d(x,y)}{max_k \left \{max_{x,y\in C_k}d(x,y)\right \} })]
           \right \}
           $
           &\\  \midrule

       7 & pbm &  $pbm$ &
           $
           \left \{
           \frac{1}{K} \times
           {\frac
                 { {\sum_{i=1}^{n}} d(x_i,y_j)}
                 {\sum_{j=1}^{N} \sum_{x_i\in C_j} d\left(x_i,y_j\right ) }}\times
           max_{i,j=1,2,..,K} d(y_i,y_j))
           \right \} $
           &\\ \midrule

       8 & I index & $Ii$&
           $
           \left \{
           [ \frac{1}{NC}\cdot \frac{\sum_{x\in D}d(x,c)} { \sum_i \sum_{x\in C_i}d(x,c_i) } \cdot max_{i,j}d(c_i,c_j) ]^P
           \right \}
           $
           &\\ \midrule

       9 & Root-mean-square std dev& $rmss$&
           $
           \left \{
           [ \frac{\sum_i\sum_{x\in C_i} \| x-c_i \|^2} {P\sum_i(n_i-1)}  ]^\frac{1}{2}
           \right \}
           $
           &\\ \midrule

       10& R-squared &$r2$&
           $
           \left \{
           1-\frac{\sum_{i}\sum_{x\in C_i}\| x-c_i \|^2}{\sum_{x\in D}\| x-c\|^2}
           \right \}
           $
           &\\ \midrule

       11& Modified Hubert $\Gamma$ statistic& $mhgs$&
           $
           \left \{
           \frac{2}{n(n-1)} \sum_{x\in D}\sum_{y\in D} d(x,y)d_{x\in C_,y\in c_j}(c_i,c_j)
           \right \}
           $
           &\\ \bottomrule

    \end{tabular}

\begin{tablenotes}
\item[1] $D$: dataset ; $n$: number of objects in $D$ ; $c$: center of $D$ ; $P$: attributes number of $D$ ; $NC$: number of clusters ; $C_i$: the i-th cluster; $n_i$: number of  objects  in $C_i$ ; $c_i$: center  of  $C_i$ ; $\sigma(C_i)$: variance vector of $C_i$ ; $d(x,y)$: distance between $x$ and $y$ ; $||X_i|| =(X_i^T\cdot X_i)^{\frac{1}{2}}$.
\end{tablenotes}
\label{tab}
\end{table*}


\begin{thebibliography}{widest-label}
\bibitem{pt}


H. E. Stanley,
Introduction to Phase Transitions and Critical Phenomena (Oxford University Press, New York, 1971).


\bibitem{qpt}
S. Sachdev, Quantum Phase Transitions, (Cambridge University Press, UK, 1999).


\bibitem{roger1}
J. Carrasquilla and  R.  Melko,
Machine learning phases of matter,
Nat. Phys. {\bf 13}, 431 (2017).

\bibitem{longising}
K. I. Aoki, T. Kobayashi,
Restricted Boltzmann Machines for the Long Range Ising Models,
Mod. Phys. Lett. B {\bf 30}, 1651401 (2016).
\bibitem{smising}
D. Kim and D. Kim,
Smallest neural network to learn the Ising criticality,
Phys. Rev. E {\bf 98}, 022138 (2018).

\bibitem{hu}
W. Hu, R. R. P. Singh, and R. T. Scalettar,
Discovering phases, phase transitions, and crossovers through unsupervised machine learning: A critical examination,
Phys. Rev. E {\bf 95}, 062122 (2017).

\bibitem{rogerxy}
M.  Beach, A. Golubeva and R. Melko,
Machine learning vortices at the Kosterlitz-Thouless transition,
Phys. Rev. B {\bf 97}, 045207 (2018).


\bibitem{Wessel}
P. Suchsland and S. Wessel, Parameter diagnostics of phases and phase transition learning by neural networks, Phys. Rev. B {\bf 97}, 174435 (2018).

\bibitem{mlpxy}
W. Zhang, J. Liu and T.-C. Wei,
Machine learning of phase transitions in the percolation and XY models,
Phys. Rev. E {\bf 99}, 032142 (2019).



\bibitem{xykt}
Kosterlitz, J. M.  Thouless and D. J. Ordering, metastability and phase transitions in two-dimensional systems, J. Phys. C {\bf 6}, 1181 (1973).
\bibitem{bkt}
Berezinskii, V. Destruction of long-range order in one-dimensional and two-dimensional systems possessing a continuous symmetry group. II. Quantum systems, Sov. J. Exp. Theor. Phys. {\bf 34}, 610 (1972).


\bibitem{dm}
V. Singh and
J. Han,
Application of machine learning to two-dimensional Dzyaloshinskii-Moriya ferromagnets,
Phys. Rev. B {\bf 99}, 174426, (2019).




\bibitem{sky}
I. A. Iakovlev,
O. M. Sotnikov and
V. V. Mazurenko
Supervised learning approach for recognizing magnetic skyrmion phases,
Phys. Rev. B {\bf  98}, 174411 (2018).


\bibitem{Potts}
C.  Li, D.  Tan and  F.  Jiang,
Applications of neural networks to the studies of phase transitions of two-dimensional Potts models, Ann. Phys.
 {\bf 391}, 312 (2018).



\bibitem{confu1}
E. van Nieuwenburg, Y. Liu and S. Huber,
Learning phase transitions by confusion,
Nat. Phys. {\bf 13}, 435 (2017).

\bibitem{fermi}
P. Broecker,
J. Carrasquilla,
R. Melko and
S. Trebst,
Machine learning quantum phases of matter beyond the fermion sign problem,
 Sci. Rep.  {\bf 7} 8823 (2017).


\bibitem{xuefeng}
X. Dong,
F. Pollmann and
X. Zhang,
Machine learning of quantum phase transitions,
Phys. Rev. B {\bf 99}, 121104 (2018).


\bibitem{wang1}
L. Wang,
Discovering Phase Transitions with Unsupervised Learning,
Phys. Rev. B {\bf 94}, 195105, (2016).

\bibitem{unp}
S. J. Wetzel,
Unsupervised learning of phase transitions: From principal component analysis to variational autoencoders,
Phys. Rev. E {\bf 96}, 022140 (2017).

\bibitem{unsup1}
K. Ch'ng,
N. Vazquez and
E. Khatami,
Unsupervised machine learning account of magnetic transitions in the Hubbard model,
Phys. Rev. E {\bf 97}, 13306 (2018).


\bibitem{tony}
T. C. Scott, M. Therani, and X. M. Wang,
Data Clustering with Quantum Mechanics,  Mathematics, {\bf 5}, 5 (2017).


\bibitem{dy}
E. van Nieuwenburg,
E. Bairey and
G. Refael,
Learning phase transitions from dynamics,
Phys. Rev. B {\bf 98}, 060301 (2018).





 \bibitem{nonequi2}
J. Venderley, V. Khemani and  E. Kim, Machine learning out-of-equilibrium phases of matter,
Phys. Rev. Lett. {\bf 120}, 257204 (2018).


\bibitem{out2}
C. Casert, T. Vieijra,
J. Nys and
J. Ryckebusch,
Interpretable machine learning for inferring the phase boundaries in a nonequilibrium system,
Phys. Rev. E {\bf  99}, 023304 (2019).



\bibitem{adv}
P. Huembeli,
A. Dauphin and
P. Wittek,
Identifying quantum phase transitions with adversarial neural networks,
Phys. Rev. B {\bf 97}, 134109 (2018).


\bibitem{dcn}
Y.  Liu and
E. P. L. van Nieuwenburg,
Discriminative Cooperative Networks for Detecting Phase Transitions,
Phys. Rev. Lett. {\bf 120}, 176401 (2018).


\bibitem{rs}
S. Efthymiou, M. J. S. Beach, and R. G. Melko,
Super-resolving the Ising model with convolutional neural networks,
Phys. Rev. B {\bf 99}, 075113 (2019).


\bibitem{ex1}
W. Lian et al.
Machine Learning Topological Phases with a Solid-State Quantum Simulator,
Phys. Rev. Lett. {\bf 122}, 210503 (2019).

\bibitem{ex2}
A.  Bohrdt et al.,
Classifying snapshots of the doped Hubbard model with machine learning,
Nat. Phys. {\bf 15}, 921 (2019).


\bibitem{topo-inv}
P. Zhang, H. Shen and H. Zhai,
Machine Learning Topological Invariants with Neural Networks,
Phys. Rev. Lett. {\bf 120}, 066401 (2017).



\bibitem{z2spin}
Y. Zhang,
R. Melko and
E. Kim,
Machine learning Z2 quantum spin liquids with quasiparticle statistics,
Phys. Rev. B {\bf 96}, 245119 (2017).

\bibitem{tp2}
Y. Zhang and  E. A. Kim,
Quantum Loop Topography for Machine Learning,
Phys. Rev. Lett. {\bf 118}, 216401 (2017).

\bibitem{mbl}
Y. Hsu,
X. Li,
D. Deng and
S. Das Sarma,
Machine Learning Many-Body Localization: Search for the Elusive Nonergodic Metal,
Phys. Rev. Lett. {\bf 121}, 245701 (2018).

\bibitem{jgliu}
Z. Cai and J. Liu
Approximating quantum many-body wave functions using artificial neural networks,
Phys. Rev. B {\bf 97},035116 (2018).

\bibitem{ml}
R. Vargas-Hern\'andez,
J. Sous,
M. Berciu and
R. Krems,
Extrapolating Quantum Observables with Machine Learning: Inferring Multiple Phase Transitions from Properties of a Single Phase,
Phys. Rev. Lett. {\bf 121}, 255702 (2018).


\bibitem{selfmap}
A. Shirinyan,
V.  Kozin,
J. Hellsvik and
M. Pereiro,
Self-organizing maps as a method for detecting phase transitions and phase identification,
Phys. Rev. B {\bf 99}, 041108 (2019).



\bibitem{tool}
C. Giannetti,
B. Lucini and
D. Vadacchino,
Machine Learning as a universal tool for quantitative investigations of phase transitions,
 Nucl. Phys. {\bf 944} 114639 (2019).


\bibitem{spinor}
J. Greitemann,
K. Liu and
L. Pollet
Probing hidden spin order with interpretable machine learning,
Phys. Rev. B {\bf 99}, 060404 (2019)


\bibitem{conf}
S. S. Lee and B. J. Kim,
Confusion scheme in machine learning detects double phase transitions and quasi-long-range order,
Phys. Rev. E {\bf  99}, 043308 (2019). % percoltion and xy model

\bibitem{order}
P. Ponte and R. G. Melko,
Kernel methods for interpretable machine learning of order parameters,
Phys. Rev. B {\bf 96}, 205146 (2017).



\bibitem{exp}
Z. Li, M. Luo, and X. Wan,
Extracting critical exponents by finite-size scaling with convolutional neural networks, Phys. Rev. B {\bf 99}, 075418 (2019).



\bibitem{rev}
Giuseppe Carleo et al., Machine learning and the physical sciences,
Rev. Mod. Phys. {\bf 91}, 045002 (2019).

\bibitem{nature}
J. F. Rodriguez-Nieva and  M. S. Scheurer,
Identifying topological order through unsupervised machine learning,
Nat. Phys. {\bf 15}, 790 (2019).

\bibitem{lafon}
R. R Coifman, S. Lafon, A. B. Lee, M. Maggioni, B. Nadler, F.  Warner, S.W. Zucker. Geometric diffusions as a tool for harmonic analysis and structure definition of data: Diffusion maps, PNAS. {\bf 102} (21), 7426(2005).

\bibitem{CH}
T. Cali\'nski and  J. Harabasz,
A dendrite method for cluster analysis,
Commun. Stat. - Theory Methods
{\bf 3}, (1974).

\bibitem{lunkuo}
P. J. Rousseeuw,
Silhouettes: a graphical aid to the interpretation and validation of cluster analysis,
J. Comput.  Appl. Math. {\bf 20}, 53 (1987).

\bibitem{SW}
R. H. Swendsen, J. S. Wang, Nonuniversal critical dynamics in Monte Carlo simulations, Phys. Rev. Lett. {\bf 58} 86 (1987).

\bibitem{prep}
J. F. Rodriguez-Nieva and  M. S. Scheurer, arXiv:1805.05961v1.


\bibitem{11index}
Y. Liu, Z. Li, H. Xiong, X. Gao, J. Wu and S. Wu, Understanding and Enhancement of Internal Clustering Validation Measures, in IEEE Transactions on Cybernetics, {\bf 43},  3,   (2013).


\bibitem{metro}
N. Metropolis, A.W. Rosenbluth, M.N. Rosenbluth, A.H. Teller, E. Teller,
Equation of State Calculations by Fast Computing Machines, J. Chem.
Phys. {\bf 21} 1087 (1953).

 \bibitem{wf}
  U. Wolff, Collective Monte Carlo Updating for Spin Systems, Phys. Rev. Lett. {\bf 62}, 361 (1989).

\bibitem{kao}
Y. Hsieh, Y. J. Kao and A. W. Sandvik.
Finite-size scaling method for the Berezinskii-Kosterlitz-Thouless transition, J Stat. Mech-Theory {\bf 2013}, P09001 (2013).

\bibitem{0.7}
B. Nienhuis, Exact Critical Point and Critical Exponents
of O($n$) Models in Two Dimensions, Phys. Rev. Lett.
\textbf{49}, 1062 (1982); Y. Deng, T. Garoni, W. Guo, H. Bl\"ote
and A. Sokal, Cluster simulations of loop models on two-
dimensional lattices, Phys. Rev. Lett. \textbf{98}, 120601 (2007).

\bibitem{q2mc}
D. M. H\"ubscher and S. Wessel, Stiffness jump in the generalized XY model on the square lattice, Phys. Rev. E {\bf 87}, 062112 (2013).


%\bibitem{q3mc}
%
%Kosterlitz-Thouless and Potts transitions in a generalized XY model
%
%Authors: Gabriel A. Canova, Yan Levin, Jeferson J. Arenzon
%
%
%
%Journal ref: Phys. Rev. E 89, 012126 (2014)
%\bibitem{pot_wu}
%RMP 2,010 citations
%The Potts model
%F. Y. Wu
%Rev. Mod. Phys. 54, 235 (1982) - Published 1 January 1982

\bibitem{q8mc}
G. A. Canova, Y. Levin, and J. J. Arenzon, Competing nematic interactions in a generalized XY model in two and three dimensions, Phys. Rev. E {\bf 94}, 032140 (2016).

\bibitem{gp}
F. Pedregosa, G. Varoquaux, A. Gramfort et. al.,
Scikit-learn: Machine Learning in Python,
Journal of Machine Learning Research,
{\bf 12}, 2825 (2011).



%doi: 10.1109/TSMCB.2012.2220543

\bibitem{localsigma}
L. Zelnik-Manor and P. Perona, Self-Tuning Spectral Clustering,
Neural Inf. Process. Syst. {\bf17} 1601 (2004);
G. Mishne and I. Cohen,
Multiscale Anomaly Detection Using DFs,
IEEE Journal of Selected Topics in Signal Processing (JSTSP), {\bf 7}, 111 (2013).

\bibitem{band}
M. S. Scheurer and  Robert-Jan Slager,
Unsupervised machine learning and band topology,
Phys. Rev. Lett. {\bf 124}, 226401 (2020).
\bibitem{dff}
Y. Long, J. Ren and H. Chen,
Unsupervised Manifold Clustering of Topological Phononics, Phys. Rev. Lett.
{\bf 124}, 185501 (2020).





\end{thebibliography}
\end{document}